\documentclass[iop, aas_macros]{emulateapj}

\usepackage{graphicx}
\usepackage{amssymb}
\usepackage{amsmath}
\usepackage{natbib}
\usepackage{epstopdf}
\slugcomment{}

\shorttitle{Exocometary CO in the Fomalhaut belt}
\shortauthors{Matr\`a et al.}

\begin{document}


\title{Detection of exocometary CO within the 440 Myr-old Fomalhaut belt: \\ a similar CO+CO$_2$ ice abundance in exocomets and Solar System comets}


\author{L. Matr\`a\altaffilmark{1}, M. A. MacGregor\altaffilmark{2}, P. Kalas\altaffilmark{3, 4}, M. C. Wyatt\altaffilmark{1}, G. M. Kennedy\altaffilmark{1}, D. J. Wilner\altaffilmark{2}, G. Duchene\altaffilmark{3, 5}, A. M. Hughes\altaffilmark{6}, M. Pan\altaffilmark{7}, A. Shannon\altaffilmark{8,9,1}, M.\ Clampin\altaffilmark{10}, M. P. Fitzgerald\altaffilmark{11}, J. R. Graham\altaffilmark{3}, W. S. Holland\altaffilmark{12}, O. Pani\'c\altaffilmark{13}, K. Y. L. Su\altaffilmark{14}}

\altaffiltext{1}{Institute of Astronomy, University of Cambridge, Madingley Road, Cambridge CB3 0HA, UK}
\altaffiltext{2}{Harvard-Smithsonian Center for Astrophysics, 60 Garden Street, Cambridge, MA 02138, USA}
\altaffiltext{3}{Astronomy Department, University of California, Berkeley CA 94720-3411, USA}
\altaffiltext{4}{SETI Institute, Mountain View, CA, 94043, USA}
\altaffiltext{5}{Univ. Grenoble Alpes/CNRS, IPAG, F-38000 Grenoble, France}
\altaffiltext{6}{Department of Astronomy, Van Vleck Observatory, Wesleyan University, 96 Foss Hill Dr., Middletown, CT 06459, USA}
\altaffiltext{7}{MIT Department of Earth, Atmospheric, and Planetary Sciences, Cambridge, MA 02139}
\altaffiltext{8}{Department of Astronomy \& Astrophysics, The Pennsylvania State University, State College, PA 16801, USA}
\altaffiltext{9}{Center for Exoplanets and Habitable Worlds, The Pennsylvania State University, State College, PA 16802, USA}
\altaffiltext{10}{NASA Goddard Space Flight Center, Greenbelt, MD 20771, USA}
\altaffiltext{11}{Department of Physics \& Astronomy, University of California, Los Angeles, CA 90095, USA}
\altaffiltext{12}{Astronomy Technology Centre, Royal Observatory Edinburgh, Blackford Hill, Edinburgh EH9 3HJ, UK}
\altaffiltext{13}{School of Physics and Astronomy, University of Leeds, Woodhouse Lane, Leeds, LS2 9JT, UK}
\altaffiltext{14}{Steward Observatory, University of Arizona, 933 North Cherry Avenue, Tucson, AZ 85721, USA}

\email{l.matra@ast.cam.ac.uk}






\begin{abstract}
Recent ALMA observations present mounting evidence for the presence of exocometary gas released within Kuiper belt analogues around nearby main sequence stars. This represents a unique opportunity to study their ice reservoir at the younger ages when volatile delivery to planets is most likely to occur. We here present the detection of CO J=2-1 emission co-located with dust emission from the cometary belt in the 440 Myr-old Fomalhaut system. Through spectro-spatial filtering, we achieve a 5.4$\sigma$ detection and determine that the ring's sky-projected rotation axis matches that of the star. The CO mass derived ($0.65-42 \times10^{-7}$ M$_{\oplus}$) is the lowest of any circumstellar disk detected to date, and must be of exocometary origin. Using a steady state model, we estimate the CO+CO$_2$ mass fraction of exocomets around Fomalhaut to be between 4.6-76\%, consistent with Solar System comets and the two other belts known to host exocometary gas. This is the first indication of a similarity in cometary compositions across planetary systems that may be linked to their formation scenario and is consistent with direct ISM inheritance.  In addition, we find tentative evidence that $(49\pm 27)$\% of the detected flux originates from a region near the eccentric belt's pericentre. If confirmed, the latter may be explained through a recent impact event or CO pericentre glow due to exocometary release within a steady state collisional cascade. In the latter scenario, we show how the azimuthal dependence of the CO release rate leads to asymmetries in gas observations of eccentric exocometary belts.
\end{abstract}






\section{Introduction}
\label{sect:intro}
Icy comets originating from the Kuiper belt or the Oort cloud are believed to be the most pristine bodies in our own Solar System, relics of the environment where the planets formed and evolved \citep[see][and references therein]{Mumma2011}. 
The unambiguous detection of exocometary volatiles in extrasolar debris disks, young Kuiper belt analogues around main-sequence stars, has recently given us the opportunity to expand such compositional studies to planetary systems beyond our own. Crucially, these volatile-bearing systems are observed at a very dynamically active phase of evolution (ages of $\sim$tens of Myr), when terrestrial planet formation is at its final stages and volatile delivery is most likely to take place \citep[][and references therein]{Morbidelli2012}. Moreover, comparison of volatile compositions in systems across a range of ages and host star properties may yield important clues on the formation of these belts within the protoplanetary disk \citep[for a review, see][]{Wyatt2015}.

Compositional studies of volatiles in exocometary belts have been carried out in terms of elemental abundances from observations of daughter atomic species \citep[as done in $\beta$ Pictoris, e.g.][]{Roberge2006, Brandeker2016}. In addition, metallic gas originating from sublimation or collisional evaporation of refractory elements has been detected and characterised in a few systems \citep[e.g.][]{Redfield2007,Nilsson2012, Hales2017}. More recently, we have also been able to use observations of CO gas emission at millimetre wavelengths, where this is likely produced either as a parent molecule or through the photodestruction of CO$_2$ \citep{Dent2014, Matra2015, Marino2016, Matra2017}.
 While detection of exocometary parent molecules is the most direct route to the composition of exocometary ices, it is challenging due to generally short survival timescale of gas molecules against stellar and interstellar UV photodissociation. This means that detection (so far limited to the CO molecule) has only been achievable with the extreme increase in sensitivity brought by the Atacama Large Millimeter/submillimeter Array (ALMA). On the other hand, atomic daughter products from molecular photodestruction are long-lasting \citep[e.g.][]{Fernandez2006, Brandeker2011, Kral2017} and may viscously expand over time to form an atom-dominated accretion disk, a picture that is consistent with all far infrared and millimetre observations of exocometary gas in the $\beta$ Pictoris system to date \citep{Kral2016}.
 
 The nearby \cite[7.7 pc,][]{vanLeeuwen2007}, 440$\pm$40 Myr-old \citep{Mamajek2012}, A3V \citep{Gray1989} star Fomalhaut is the 18th brightest star at visible wavelengths beyond our own Sun. The star hosts a planetesimal belt producing dust first detected through its excess emission above the stellar photosphere at infrared wavelengths \citep[e.g.][]{Aumann1985} and later imaged in thermal \citep{Holland1998, Holland2003} and scattered \citep{Kalas2005, Kalas2013} light. These revealed that the belt is confined to a ring of $\sim15$ AU width at a distance of $\sim$140 AU from the star, and that it has a significant eccentricity of $\sim$0.1. The latter causes pericentre glow at infrared wavelengths, due to material being significantly closer to the central star with respect to apocentre, causing it to be hotter and brighter \citep{Wyatt1999, Stapelfeldt2004, Marsh2005, Acke2012}. 
 This eccentricity and the sharpness of the belt's inner edge were attributed to a shepherding planet-mass companion \citep{Kalas2005}.  The subsequent \textit{Hubble Space Telescope} (HST) discovery of the very low mass companion Fomalhaut b \citep{Kalas2008} appeared consistent with the hypothesis that a planet could sculpt the inner edge \citep{Quillen2006, Chiang2009}, until further observations showed that Fomalhaut b's orbit is highly eccentric \citep{Kalas2013, Beust2014, Pearce2015}.  The existence of Fomalhaut b has been independently replicated \citep{Currie2012,Galicher2013}, but its physical properties continue to be investigated \citep{Marengo2009, Kennedy2011, Janson2012, Tamayo2014, Kenyon2014, Neuhauser2015, Kenyon2015, Lawler2015, Janson2015}.

ALMA 850 $\mu$m imaging of part of the belt (near the NW ansa) confirmed the steepness of the inner and outer edge of the parent body distribution \citep{Boley2012}. Further, higher resolution 1.3 mm imaging \citep{White2017}, though only covering the region along the ring's minor axis, tightened the constraints on the ring's width, as well as confirming the slope of the size distribution previously obtained through Australian Telescope Compact Array (ATCA) 7 mm observations \citep{Ricci2012}.
However, only complete millimetre imaging of the ring can constrain its azimuthal morphology and eccentricity. We achieved the latter through a new 1.3 mm mosaic obtained with ALMA and described in \citet{Macgregor2017}; in particular, we confirm that the geometry of the parent body ring resembles that of the smaller grains, and achieve the first conclusive detection of apocentre glow, caused by the higher surface density expected at apocentre with respect to pericentre, due to particles on eccentric orbits spending more time at apocentre \citep{Pan2016}. 

In parallel to the latest observations of the dust ring, recent deep searches for molecular CO J=3-2 emission around the belt's NW ansa at 345 GHz with ALMA \citep{Matra2015}, as well as searches for atomic ionised carbon (CII) and neutral oxygen (OI) through far-infrared \textit{Herschel} spectroscopy \citep{Cataldi2015} yielded null results. The CO non-detection was used to set an upper limit to the CO ice content in the planetesimals, but this was still consistent with (and close to the upper boundary of) the range of CO abundances observed in Solar System comets. On the other hand the non-detection of atomic gas, tracing the bulk of the gas in an exocometary origin scenario, was used to derive a low upper limit to the gas/dust ratio; being well below 1, this ruled out gas-dust interactions as the origin for the narrow eccentric ring in the Fomalhaut system.

In this work, we present deeper ALMA CO J=2-1 observations of the entirety of the Fomalhaut belt, yielding the first detection of gas in the system. In Sect. \ref{sect:obs} we describe the observations, focusing on the CO imaging procedure. A description of the detection through the spectro-spatial filtering technique first applied in \citet{Matra2015} is in Sect. \ref{sect:spatfilt} and \ref{sect:spectrospatfilt}, followed by the constraints this detection sets on the ring's rotation axis (Sect. \ref{sect:spectrospatfilt}), total CO mass (Sect. \ref{sect:totflux}), and the consistency of the detection with the previous observation of the CO J=3-2 transition (Sect. \ref{sect:B7check}). Finally, we analyse the radial and azimuthal morphology of the CO emission (Sect. \ref{sect:morph}). 

We then go on to discuss the implications of this result, including proving the exocometary origin of the gas (Sect. \ref{sect:exoorig}, \ref{sect:stoch}, \ref{sect:stst}), drawing the first comparison between the ice content of extrasolar and Solar System comets (Sect. \ref{sect:comparison}), and discussing the possible origin of the observed similarity with a particular emphasis on ISM volatile inheritance (Sect. \ref{sect:similorigin}). At the same time, we investigate the possible cause of the tentative CO enhancement observed at the belt's pericentre, focusing on a general prediction of CO pericentre or apocentre glow in eccentric exocometary belts such as Fomalhaut (Sect. \ref{sect:COperiglow}).
We conclude by summarising the outcomes of this study in Sect. \ref{sect:concl}.

\section{Observations}
\label{sect:obs}

We observed the Fomalhaut belt with the ALMA array using Band 6 receivers, allowing simultaneously a wide bandwidth for 1.3mm dust continuum imaging and a native spectral resolution of 1.27 km/s around the rest frequency of the CO J=2-1 transition (230.538 GHz). The observational setup, as well as the calibration, data reduction and continuum imaging process are described in detail in \citet{Macgregor2017}.  Within the spectral window containing the CO line, we carried out continuum subtraction on the combined visibility dataset using the \texttt{uvcontsub} task within the \texttt{CASA} software package \citep[version 4.5.2,][]{McMullin2007}. We image the visibility dataset using the \texttt{CLEAN} task.  We do not apply any image deconvolution, but carry out our analysis (next Section) on the primary beam-corrected dirty CO data cube. To enhance detectability of low CO surface brightness, we apply a taper within \texttt{CLEAN} to achieve an imaging resolution that better matches the width of the ring ($2\farcs36\times2\farcs13$, as opposed to the $1\farcs52\times1\farcs12$ resolution achieved from natural visibility weights).  These imaging steps should not affect our results significantly compared to using the visibilities directly, since our u-v coverage yields a well-behaved dirty beam with little sidelobe emission, never above the $\sim$4\% level. Our final data cube has a root-mean-square (RMS) noise level of 0.7 mJy beam$^{-1}$ in a 1.3 km/s channel for a beam size of $18.2\times16.4$ AU at the distance of the Fomalhaut system. The expected systematic flux calibration accuracy is of order $\sim10$\% \citep{Fomalont2014}.

\section{Results and Analysis}
\label{sect:res}

\begin{figure*}
  \includegraphics*[scale=0.6]{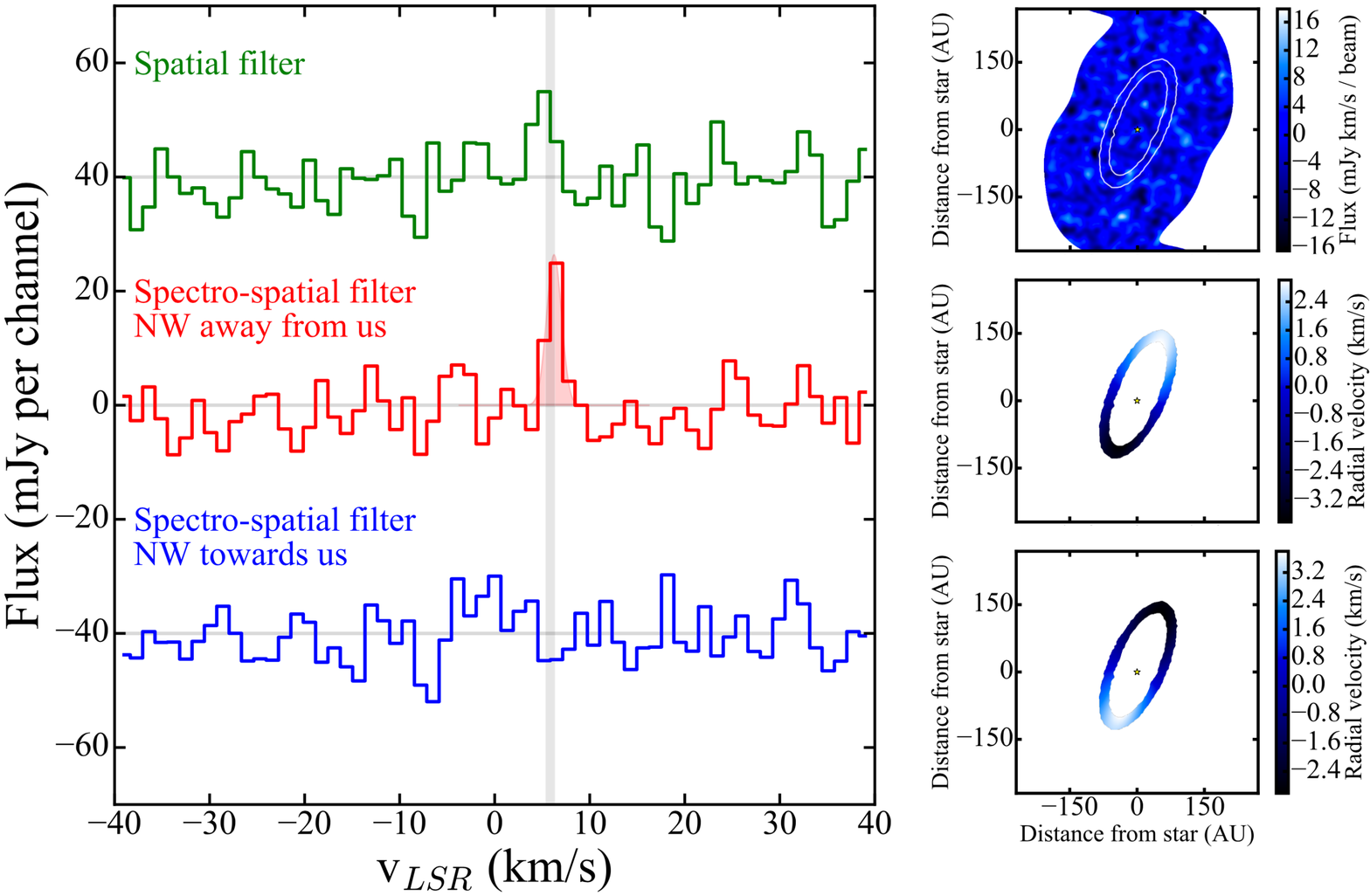}
\caption{CO J=2-1 ($\nu_{\rm rest} = 230.538 $ GHz) spectra of the Fomalhaut system obtained using different filtering techniques to achieve maximum signal-to-noise ratio (SNR). \textit{Top:} Spectrum obtained through a spatial filter, i.e. by spatially integrating emission over the area where the dust continuum from the ring is detected at a level $>4\sigma$ (shown by contours overlaid on the moment-0 image displayed on the right).
\textit{Centre:} Spectrum obtained by applying a spectro-spatial filter. The same spatial filter as above is applied after shifting the 1D spectrum at each spatial location by an amount equal to the opposite of its expected Keplerian radial velocity (see corresponding velocity map on the right). 
\textit{Bottom:} Spectrum obtained by applying the same spectro-spatial filter, but assuming a reversed velocity field, i.e. with the NW ansa coming towards us (again, see corresponding velocity map on the right). A $3.3\sigma$ peak at a velocity consistent with that of the star (grey shaded region) is obtained using the spatial filter (top), and is boosted to $5.4\sigma$ using the spectral filter with the NW ansa moving away from us (centre). The red shaded region is the spectrally unresolved Gaussian that best fits the filtered data.}
\label{fig:spectra}
\end{figure*}

\subsection{Spatial filter: CO J=2-1 detection}
\label{sect:spatfilt}
Fig. \ref{fig:spectra} (top right) shows the CO J=2-1 moment-0 map of the Fomalhaut ring, spectrally integrated within $\pm3.5$ km/s around the expected radial velocity of the star \citep[$v_{\star, \rm LSR}=5.8\pm0.5$ km/s][]{Gontcharov2006}. No statistically significant emission is observed above a noise distribution that is well approximated by a Gaussian, and that has a RMS of 4 mJy km s$^{-1}$ beam$^{-1}$. The few 3-4$\sigma$ peaks observed are in line with the expectation from the Gaussian noise distribution and the large number of resolution elements in the image. The contour lines indicate the sky region where continuum emission is detected above the 4$\sigma$ level \citep{Macgregor2017}, corresponding to an area comprising a number of beams of N$_{\rm beams}\sim40$. 
As done for previous Band 7 CO observations of the Fomalhaut system \citep{Matra2015}, we now proceed by assuming that any CO present in the system should be of secondary origin and co-located with the dust from which it is produced, and later on test these assumptions (Sect. \ref{sect:exoorig} and \ref{sect:morph}). This allows us to apply a spatial filter, i.e. to spatially integrate the CO data cube over this region where the continuum is detected, yielding an improvement on the signal-to-noise ratio (SNR) equal to $\sqrt{\rm N_{\rm beams}} \sim 6.4$.
The obtained 1D spectrum, shown in Fig. \ref{fig:spectra} (left), presents a 3.3$\sigma$ peak at a velocity consistent with the expected $v_{\star, \rm LSR}$. 

\subsection{Spectral + spatial filter: SNR boost and kinematics}
\label{sect:spectrospatfilt}
In order to test the robustness and improve the significance of the detection, we make the further assumption that the CO gas lies in a vertically flat disk in Keplerian rotation around a star of 1.92 $M_{\odot}$, with the same orbital parameters as the dust ring, obtained from dust continuum imaging \citep[][]{Macgregor2017} and consistent with scattered light imaging \citep[][]{Kalas2013}. Then, for each pixel location $(x_{\rm sky}, y_{\rm sky})$, where $(0,0)$ is the accurately known location of the star (detected in the continuum dataset), we can determine the expected radial velocity field $v_z (x_{\rm sky}, y_{\rm sky})$ of the gas, both for the case in which the north-west (NW) ansa is moving away from us (Fig. \ref{fig:spectra}, centre right) or towards us (Fig. \ref{fig:spectra}, bottom right). Using this predicted velocity field, we apply a spectral filter, i.e. we shift the 1D spectrum at each $(x_{\rm sky}, y_{\rm sky})$ pixel location by $-v_z (x_{\rm sky}, y_{\rm sky})$, moving any CO emission from the predicted disk velocity to the stellar velocity $v_{\star, \rm LSR}$. If we then sum together the contributions from all pixels where the dust continuum is detected (or in other words, if we apply the spatial filter), we obtain a spectrum which is free from noise originating from spectral channels as well as spatial locations where no CO emission is expected. As shown in Fig. \ref{fig:spectra} (centre left, for the velocity field in centre right), this method boosts the SNR of the peak to $5.4\sigma$. 

The spectral filter yields kinematic information on the orbiting gas, since the SNR of the CO line is maximised only for the correct velocity field. This is because for an incorrect velocity field, the spectral filter spreads CO emission over a larger number of velocity channels, reducing the SNR (see Fig. \ref{fig:spectra}, bottom left and right). This proves that CO gas in the NW ansa is moving away from us, while gas in the south-east (SE) ansa is coming towards us (Fig. \ref{fig:spectra}, centre right). This is consistent with the kinematics of the star, whose SE part was also found to be moving towards us \citep{LeBouquin2009}. We note that this information is still insufficient to derive the sense of rotation of the ring on-sky (clockwise or anti-clockwise), and in turn which side of the ring is closer to Earth and forward scattering \citep[see discussion in][]{Kalas2013, Min2010}. This is because both the star and the CO only inform us on the sky-projected rotation axis, which is perpendicular to the sky-projected major axis of the ring and points in the north-east (NE) direction. However, the crucial missing piece of information is whether the ring's rotation axis is pointing towards us (i.e. inclination of $\sim+65^{\circ}$ to the plane of the sky) or away from us ($\sim-65^{\circ}$ inclination). Therefore, the current data is insufficient to determine whether the ring is rotating in a clockwise or anticlockwise direction on-sky, and whether the brighter NE side observed by HST is in front or behind the sky plane.

\subsection{Total flux and CO mass}
\label{sect:totflux}

In order to measure the integrated CO J=2-1 line flux and the velocity centroid, we fit a Gaussian to the spectro-spatially filtered spectrum derived above and shown in Fig. \ref{fig:spectra} (centre left).  The best-fit Gaussian width is consistent with the line being unresolved, as expected from application of our filtering method, and the best-fit velocity is $6.1\pm0.2$ km/s, consistent with the expected stellar velocity of $5.8\pm0.5$ km/s. The best-fit integrated line flux is $68\pm16$ mJy km/s (including the absolute flux calibration uncertainty, added in quadrature to the uncertainty from the Gaussian fit), or $(5.2\pm1.2)\times10^{-22}$ Wm$^{-2}$. 

Assuming that the CO is optically thin to the line of sight at the observed frequency (which we verify in the next paragraph), we use a non-local thermodynamic equilibrium (NLTE) molecular excitation analysis \citep{Matra2015} to derive constraints on the total CO mass in the system. The derived mass value depends on two unknown parameters, the kinetic temperature of the gas and density of the main collisional partners. We take the main colliders to be electrons, since these are most likely to be the dominant species for which CO collision rates are known if the gas is of exocometary origin \citep{Matra2015}. We cover the full electron density parameter space between the radiation-dominated and local thermodynamic equilibrium (LTE) regimes, and probe a wide range of kinetic temperatures between 10 and 1000 K. The CO mass derived is in the range $0.65-42 \times10^{-7}$ M$_{\oplus}$, where the boundaries were obtained from the $\pm1\sigma$ limits on the integrated line flux. This is the lowest CO mass detected in any circumstellar disk to date, which is readily understood by noting that Fomalhaut, at a distance of 7.7 pc, is the nearest circumstellar disk where CO has been searched for by ALMA to date. We will discuss the implication of this mass measurement in Sect. \ref{sect:exoorig}. 

We note that the CO excitation model does not yet account for the effect of infrared or UV pumping, i.e. transitions between vibrational and electronic levels within the molecule. These are excited in the presence of a strong infrared and/or UV radiation field, as may be the case around an A star such as Fomalhaut. As molecules relax to the ground electronic and/or vibrational levels, higher rotational levels may be more populated than predicted through CMB radiation alone, affecting the rotational level populations. Nonetheless, this would only influence the molecule in the radiation-dominated regime, or in other words it would only change our upper limit on the derived range of CO masses. Since less mass would be needed to produce the same flux in the presence of significant UV/IR pumping, our derived CO mass range can be taken as a conservative estimate, and is likely narrower than derived above.

We now carry out a check to probe the optical thickness of the CO line. We assume that the CO density is uniform in a disk with inner radius at $R=136.3$ AU and $\Delta R=13.5$ AU wide, with a constant aspect ratio $h=H/R$ (height above the midplane divided by the radius) equal to the best-fit mean proper eccentricity of the planetesimals from continuum observations \citep[0.06,][]{Macgregor2017}. This way, we can derive the ring volume and hence measure an average CO number density for our range of CO masses, obtaining a range of number densities between $2-75\times10^{-2}$ cm$^{-3}$. 
We then estimate the maximum column density using the longest path length along the line of sight passing through our simple model ring described above. We neglect any column density enhancement expected at the two ansae due to projection effects (see Sect. \ref{sect:COperiglow}). For the range of number densities above, this yields maximum column densities of $0.04-1.67\times10^{14}$ cm$^{-2}$ for a maximum line-of-sight path length of $\sim$15 AU. 
Finally, we calculate the maximum optical thickness through its definition \citep[Eq. 3 in][]{Matra2017}, for the full range of electron densities and kinetic temperatures probed above, and for an intrinsic line width taken to be equal to the Doppler broadening expected from each of the temperatures considered. The maximum, worst-case scenario optical thickness we obtain is $\tau\lesssim 0.18$ (with values ranging down to 10$^{-4}$ depending on the parameters), which confirms our optically thin assumption and our CO mass measurement. 

\subsection{Consistency with archival Band 7 observations}
\label{sect:B7check}

\begin{figure}
  \includegraphics*[scale=0.43]{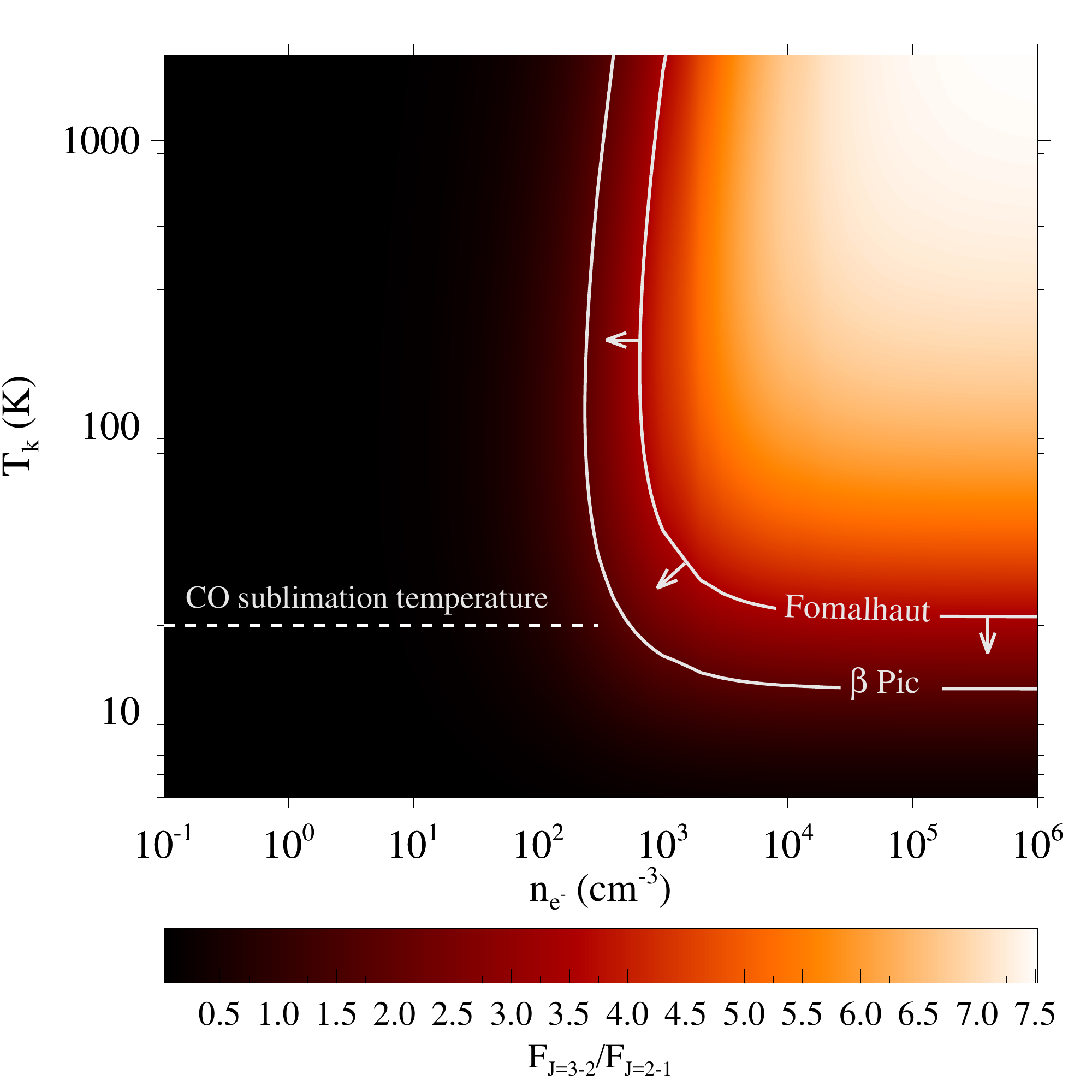}
\vspace{-4mm}
\caption{Colour map of the CO J=3-2/J=2-1 line ratios expected for a wide range of kinetic temperatures $T_{\rm k}$ and electron densities $n_{\rm e^-}$. The disk-averaged upper limit we estimate in the Fomalhaut belt is consistent with the value found for the $\beta$ Pic disk, implying that similar excitation conditions in the two disks are still possible and explaining the CO J=3-2 Fomalhaut non-detection in \citet{Matra2015}.}
\label{fig:linerat}
\end{figure}

We now check that our detection of the CO J=2-1 line is consistent with the CO J=3-2 upper limit available from previous ALMA Band 7 observations \citep{Matra2015}. For a well-characterised mm radiation field such as that around Fomalhaut \citep[see Fig. 3 in][]{Matra2015}, line ratios of optically thin molecular transitions only depend on two free parameters determining the level populations of the upper level of each transition. These are the gas kinetic temperature $T_{\rm k}$ and the density of collisional partners $n_{\rm e^-}$. Taking our J=2-1 integrated line flux of $(5.2\pm1.2)\times10^{-22}$ Wm$^{-2}$, and the 3$\sigma$ upper limit of $18\times10^{-22}$ Wm$^{-2}$ on the J=3-2 line flux from \citet{Matra2015}, we obtain an upper limit on the average 3-2/2-1 line ratio in the Fomalhaut ring of 3.5 or 6.7, depending on whether we consider our J=2-1 measurement or its 3$\sigma$ lower limit.
This line ratio upper limit of 3.5 in Fomalhaut will therefore trace a line in $T_{\rm k}$-$n_{\rm e^-}$ space (Fig. \ref{fig:linerat}), the latter two quantities being degenerate. For comparison, we show the average line ratio measured in the $\beta$ Pictoris disk \citep{Matra2017}. While strictly speaking we can only exclude 3-2/2-1 line ratios higher than 6.7, we find that the archival non-detection of the J=3-2 transition is in agreement with a $\beta$ Pictoris-like gaseous environment (average line ratio of $1.9\pm0.3$), and fully consistent with the new J=2-1 detection. While this is purely a consistency check, we remind the reader that there is no reason to assume that the electron density and/or CO excitation conditions in the Fomalhaut belt should be the same as around $\beta$ Pictoris. Furthermore, we note once again that the introduction of UV/IR pumping in our model may influence Fig. \ref{fig:linerat} by increasing the minimum possible line ratio in the radiation-dominated regime of excitation (left hand side in the figure).


\subsection{Radial and azimuthal morphology}
\label{sect:morph}

\begin{figure}
 \hspace{-8mm}
  \includegraphics*[scale=0.40]{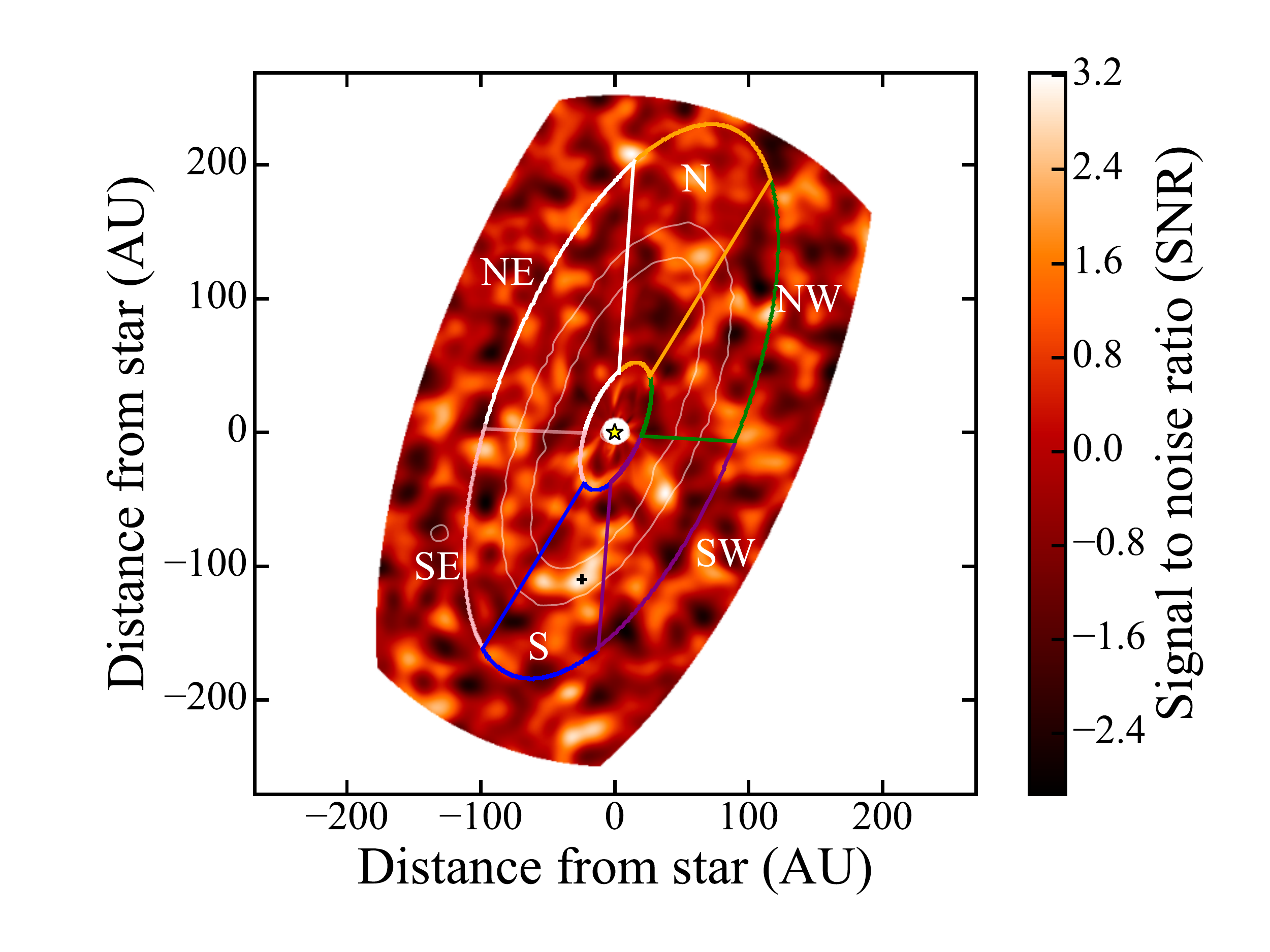}
\vspace{-8mm}
\caption{The colour scale shows the CO SNR map obtained after spectral filtering. For each pixel, given some orbital parameters (see main text) emission from the velocity channel corresponding to the expected Keplerian velocity is shown. The thin white contours represent regions where the dust continuum is detected above the 4$\sigma$ level. Wedges of different colours represent regions used to obtain the radial profiles displayed through same colours in Fig. \ref{fig:radprofs}. A significant enhancement is seen in the southern (S) wedge, and is located near the best-fit pericentre location (plus sign) as determined from ALMA dust continuum imaging \citep{Macgregor2017}.}
\label{fig:wedges}
\end{figure}

\begin{figure}
\vspace{0mm}
 \hspace{-4mm}
  \includegraphics*[scale=0.37]{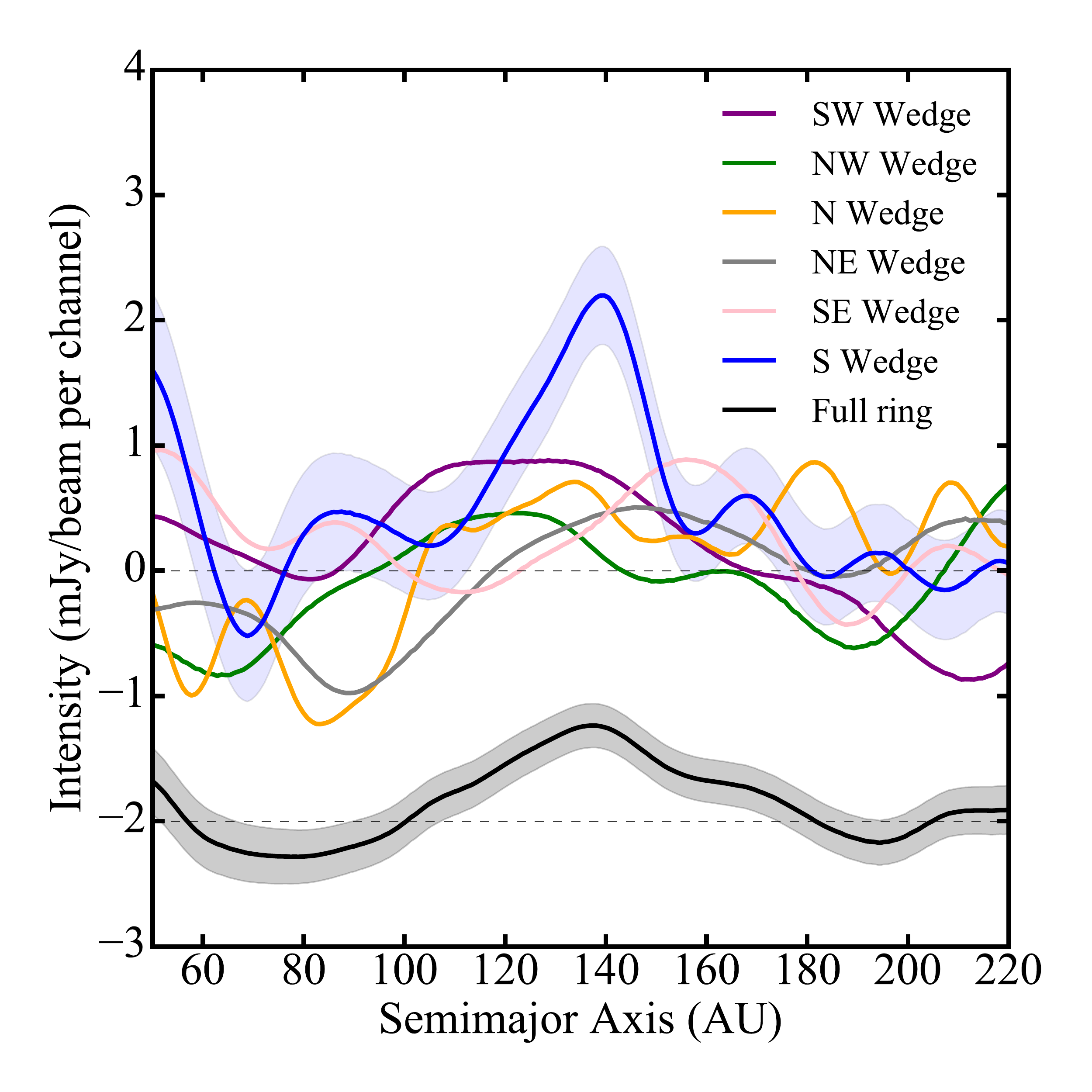}
\vspace{-8.0mm}
\caption{Radial profiles obtain through azimuthal averaging of the spectrally filtered CO map. Lines of different colours show profiles averaged within each of the wedges in Fig. \ref{fig:wedges}. The blue shaded region represent the $\pm1\sigma$ confidence interval as a function of radius within the south wedge. This interval is similar for other wedges, and not shown for clarity. The black line and grey shaded region (shifted vertically by -2 mJy beam$^{-1}$ per channel) are the same profile and confidence interval obtained through azimuthal averaging over all azimuths. }
\label{fig:radprofs}
\end{figure}

In order to extract information on the spatial distribution of CO, we relax the assumption that it must be co-located with the dust millimetre continuum. In particular, we apply the spectral filter method to the entire data cube, shifting spectra in each spatial pixel to align CO emission with the stellar velocity. Instead of spatially integrating across the area where the continuum is detected, we examine the channel map corresponding to the stellar velocity (as shown in Fig. \ref{fig:wedges}). As expected, due to spatial dilution of the emission, no obvious significant emission is seen along the whole dust ring.
For a given semimajor axis in the ring's orbital plane, taking once again the orbital parameters from dust continuum fitting, we can obtain a sky-projected orbit, and average all azimuthal contributions along it to obtain an intensity profile as a function of ring semimajor axis. The black line in Fig. \ref{fig:radprofs} shows the semimajor axis profile between 50 and 220 AU obtained by azimuthally averaging around all true anomalies, showing a CO detection at semimajor axis and width consistent with the dust continuum ring. This confirms that CO is indeed co-located with dust in the Fomalhaut belt.

Finally, we test for any indication of azimuthal asymmetry in the CO emission. To begin with, we construct the same semimajor axis profile for each of 6 azimuthal regions, or `wedges', as displayed through different colors in Fig. \ref{fig:wedges}. These profiles are shown by the corresponding colors in Fig. \ref{fig:radprofs}. For a perfectly axisymmetric on-sky surface brightness distribution of the ring, due to averaging across only $\sim$1/6$^{\rm th}$ of the azimuths, 
we expect the SNR in each wedge to decrease by a factor $\sim \sqrt{6}=2.45$ compared to the profile averaged across all azimuths, and lie at the $1-2\sigma$ level. However, the majority of CO emission appears to originate from the south (S) wedge (blue), where CO is detected at the $6.2\sigma$ level. The latter is also apparent in Fig. \ref{fig:wedges}, where we see that this region of emission is located near the ring's pericentre location as determined by ALMA dust continuum fitting \citep[$\omega=22\fdg5\pm4\fdg3$, black cross in Fig. \ref{fig:wedges},][]{Macgregor2017}. 
We find that the flux in the S wedge contributes to $(49\pm 27)$\% of the flux integrated across all azimuths; given the low SNR levels, we cannot distinguish whether all of the emission originates from this wedge or whether other regions also contain CO emission, as however hinted at by the low levels of positive emission observed across all wedges.

\begin{figure}
\vspace{0mm}
 \hspace{-5mm}
  \includegraphics*[scale=0.37]{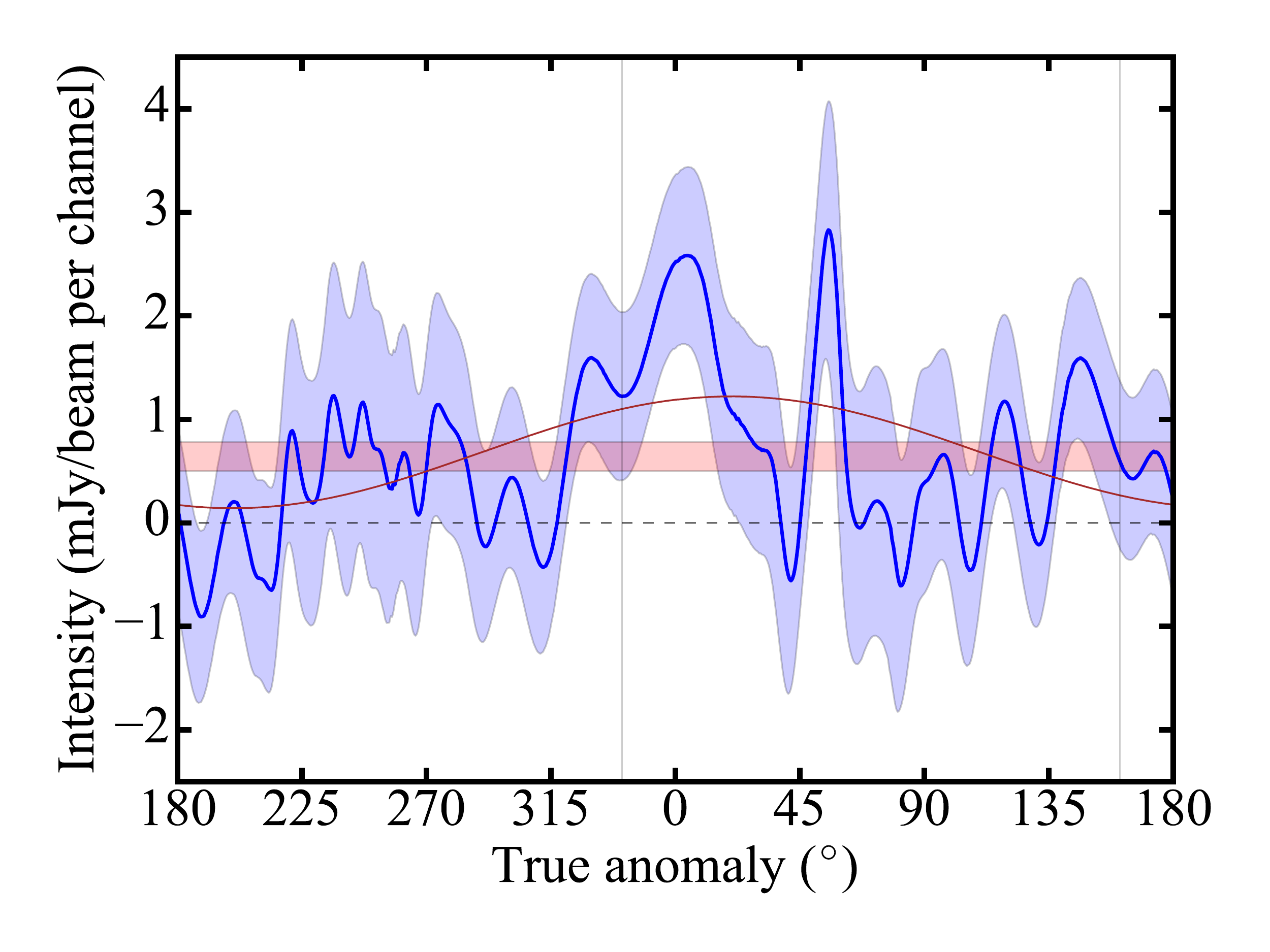}
\vspace{-8.0mm}
\caption{Radially-integrated intensity between 120 and 150 AU (blue line) as a function of true anomaly from the spectrally filtered CO map in Fig. \ref{fig:wedges}. The shaded region represents the $\pm1\sigma$ confidence interval. The red line is the cosine function that best fits the data; the red shaded region is the $\pm1\sigma$ interval of its best-fit mean, confirming our detection. The vertical thin grey lines are the locations of the on-sky ansae.}
\vspace{2mm}
\label{fig:azprof}
\end{figure}

To analyse this azimuthal variation in more detail, Fig. \ref{fig:azprof} shows the azimuthal profile obtained by integrating emission between semimajor axes of 120 and 150 AU for all true anomalies $f$. We recover the enhancement at pericentre (true anomaly $f=0^{\circ}$) and another $2.6\sigma$ peak of emission at $f\sim45^{\circ}$, also evident in Fig. \ref{fig:wedges}. We then fit a cosine function to this azimuthal profile; this is a good representation of the expectation from our steady state release model discussed in Sect. \ref{sect:stst}, for the case where a pericentre enhancement of the CO mass with respect to apocentre is predicted. We use Levenberg-Marquardt least-squares minimization to find the best-fit phase, mean intensity and pericentre to apocentre intensity enhancement ($1-I_{\rm apo}/I_{\rm peri}$) of our model cosine function. The phase obtained with respect to pericentre ($20^{\circ}\pm 25^{\circ}$) is consistent with the model having a maximum near pericentre. 
The mean ($0.68\pm 0.16$ mJy beam$^{-1}$ per 1.3 km/s channel) being significantly different from zero is further confirmation of our detection at a level consistent with the radial profile in Fig. \ref{fig:radprofs}, bottom and with the spectrum in Fig. \ref{fig:spectra}, centre left.
This cosine fit leads to an estimate a pericentre enhancement with respect to apocentre of $(88\pm25)$\%. 
We will compare such enhancement to model predictions in Sect. \ref{sect:COperiglow}.

\section{Discussion}
\label{sect:disc}
Through spectral and spatial filtering applied to new ALMA observations, we detected low levels of CO J=2-1 emission co-located with dust in the Fomalhaut ring, and measured the CO gas mass to be between 0.65 and 42 $\times10^{-7}$ M$_{\oplus}$. We then used line velocities to show that the ring's sky-projected rotation axis points to the NE, matching that of the star; in other words, the SE ansa is moving towards us, and the NW away from us. Finally, we presented tentative evidence that most of the detected emission originates near the ring's pericentre, as determined by our fitting of the ALMA dust continuum image \citep{Macgregor2017}. 

In this section, we investigate the origin of the CO observed, proving its exocometary nature, and discussing its origin in either a recent impact or steady state release. Then, we use a steady state collisional cascade model to derive exocometary ice compositions. This allows us to compare Fomalhaut with other planetary systems including the Solar System, and to make the prediction of CO pericentre or apocentre glow in eccentric exocometary rings.

\subsection{The exocometary nature of the gas}
\label{sect:exoorig}

The origin of gas remains to be found for most of the known gas-bearing debris disks. A primordial versus secondary origin dichotomy has emerged in the past years, due to the youth ($\lesssim40$ Myr) of the detected systems (see Sect. \ref{sect:intro}). Aside from a tentative detection of CO emission in the 1-2 Gyr old $\eta$ Corvi system \citep{Marino2017}, which is not co-located with the outer dust belt and remains to be confirmed, Fomalhaut is the most evolved debris belt to host CO gas, at an age of ($440\pm40$) Myr. 

Can the observed CO have survived since the protoplanetary phase of evolution? CO survival requires shielding from the interstellar UV radiation field, which otherwise rapidly photodissociates the molecule on a timescale $t_{\rm phd}$ of 120 years. This shielding can be produced by CO itself and other molecules such as H$_2$, where the latter dominates the gas mass in the primordial origin scenario. In Sect. \ref{sect:totflux}, through a simplified model, we estimated an average number density of CO in the ring of order $2-75\times10^{-2}$ cm$^{-3}$. In this model, a CO molecule sitting in the radial and vertical centre of the ring will therefore have a CO column density of order $10^{12}-10^{14}$ cm$^{-2}$ surrounding it, and assuming a low CO/H$_2$ abundance ratio of $10^{-6}$ \citep[similar to that found in the old TW Hya protoplanetary disk,][]{Favre2013}, an H$_2$ column density of order $10^{18}-10^{20}$ cm$^{-2}$. Since these CO and H$_2$ column densities are insufficient to shield CO over the system's lifetime \citep{Visser2009}, we conclude that the observed CO cannot have survived since the protoplanetary phase. We note that freeze-out onto grains is also negligible, due to the relatively low density of grains in the ring and their temperature being much above the CO freeze-out temperature \citep{Matra2015}. The observed CO must therefore be of secondary origin, i.e. recently produced through either continuous, steady state replenishment or a recent stochastic event. We analyse both possibilities below.

\subsection{Origin of the gas: stochastic collision}
\label{sect:stoch}
For a stochastic collision, the requirement is recent production of the observed CO mass, i.e. at least $6.5 \times10^{-8}$ M$_{\oplus}$. Assuming a 10\% CO+CO$_2$ mass fraction, this requires the destruction of a comet of total mass $6.5 \times10^{-7}$ M$_{\oplus}$, or about 300 Hale-Bopp masses \citep[e.g.][]{Weissman2007}. %
Given the observationally well constrained mass loss rate of small grains through the collisional cascade (Appendix \ref{sect:appb}) and the known CO mass and photodestruction rate (previous Section), we can estimate (see next Section) the collisional mass loss rate of such large, CO+CO$_2$-rich bodies in the cascade, obtaining a range between $0.012-0.046$ M$_{\oplus}$ Myr$^{-1}$. Then, we can estimate the timescale for collisions between any such supercomets to take place is 14-54 years, meaning that
the rate is 2.2-8.6 every 120 years (the survival timescale of CO). This would suggest that it is possible that the recent destruction of a large body within the belt, alone, produced all of the observed CO mass.

However, this conclusion is subject to 
the assumption that bodies of the required $6.5 \times10^{-7}$ M$_{\oplus}$ mass for CO release, which would be $\sim$100 km in size, lose mass at the same rate as other bodies participating in the collisional cascade. This implicitly assumes an extrapolation of the size distribution from small, observable grains up to bodies of this size. Although $\sim$100 km is consistent with observations of the Kuiper and asteroid belts in the Solar System \citep{Bottke2005,Fraser2014}, as well as other planetesimal growth models \citep{Johansen2015, Shannon2016}, it remains to be determined whether such large bodies in the Fomalhaut belt exist and participate in the collisional cascade. If we extrapolate the power law size distribution of \citet{Wyatt2002} to large sizes ($n(D)\propto D^{2-3q}$ with $q=11/6$, Fig. 8 in their paper), we find that there should be 7$\times$10$^7$ objects with this or larger mass within the Fomalhaut ring. Then, the collision timescale of one supercomet is $\sim1.0-3.8$ Gyr, which is longer than the $\sim$440 Myr age of the system. This indicates supercomets would have seldom collided over the age of the system, and their size distribution would therefore have been set by growth processes during planet formation, where the latter is completely unconstrained observationally.
In addition, if the extrapolation of the size distribution to these sizes were valid, the total mass of the belt ($\sim$0.4 Jupiter masses) would be $\sim$4 times higher than the 29 M$_{\oplus}$ that an initial minimum mass solar nebula (MMSN) planetesimal disk would have had between 120 and 150 AU \citep{Kenyon2008}.
In general, these high belt masses required challenge the validity of our extrapolation, and point to a likely steeper size distribution for the large primordial bodies \citep[as is the case for Kuiper belt objects, e.g.][]{Schlichting2013}.

Overall, this caveat would make our estimated high collision rate for large bodies an upper limit. To conclude, it is therefore possible that the destruction of a large icy body released all the CO observed in the Fomalhaut ring, although a reasonable likelihood for this event to happen requires the Fomalhaut belt to be massive, of order a few times higher than the expectation from a MMSN-like disk.
%



\subsection{Origin of the gas: steady state release\\ and ice composition}
\label{sect:stst}
On the other hand, we can consider the total gas release expected from the steady state collisional cascade, in the framework described in \citet{Matra2015}, and already applied to other debris disks hosting secondary gas (see Sect. \ref{sect:comparison}).
Regardless of the details of the ice removal mechanism, this method can be used to estimate the CO+CO$_2$ ice mass fraction in Fomalhaut's exocomets that is required to produce the observed CO gas.
In summary, we assume that a steady state collisional cascade is in place within the ring, with large parent bodies grinding down to produce dust of sizes all the way down to the blow-out limit \citep[e.g.][]{Wyatt2002}. 

Solid mass is being inputted through catastrophic collisions of the largest comets in the collisional cascade, and the CO+CO$_2$ mass fraction will be lost through gas release within the cascade. 
The condition of steady state imposes that the rate at which mass is being inputted by the largest bodies ($\dot{M}_{D_{\rm max}}$) is equal to the sum of the rate at which mass is being lost through CO+CO$_2$ outgassing ($\dot{M}_{\rm CO+CO_2}$) and the rate at which mass is being lost through radiation forces at the bottom of the cascade, $\dot{M}_{D_{\rm min}}$. Assuming that all of the CO+CO$_2$ ice is lost through the cascade before reaching the smallest sized grains (see discussion in next Section), we have $\dot{M}_{\rm CO+CO_2}=f_{\rm CO+CO_2}\dot{M}_{D_{\rm max}}$, meaning that we can measure the CO+CO$_2$ ice mass fraction in exocomets $f_{\rm CO+CO_2}$ through
\begin{equation}
f_{\rm CO+CO_2}=\frac{1}{1+\dot{M}_{D_{\rm min}}/\dot{M}_{\rm CO+CO_2}}.
\end{equation}

In steady state, the outgassing rate of CO+CO$_2$ molecules equals their destruction rate (where the latter is known) through $\dot{M}_{\rm CO+CO_2}=M_{\rm CO_{\rm obs}}/t_{\rm phd}$. 
Additionally, the mass loss rate $\dot{M}_{D_{\rm min}}$ of CO+CO$_2$-free particles at the small size end of the collisional cascade can be estimated by considering the collision timescale of particles just above the minimum size in the distribution (Appendix \ref{sect:appb}). The latter is well constrained by observations, and can be calculated as shown in Eq. \ref{eq:mlossratefinal}.

Overall, the CO+CO$_2$ ice mass fraction of exocomets can therefore be estimated through
\begin{equation}
f_{\rm CO+CO_2}=\frac{1}{1+0.0012 \ R^{1.5}\Delta R^{-1}f^2L_{\star}M_{\star}^{-0.5}t_{\rm phd}M_{\rm CO_{\rm obs}}^{-1}},
\end{equation}
where $R$ and $\Delta R$ are in AU, $L_{\star}$ and $M_{\star}$ are in L$_{\odot}$ and M$_{\odot}$, $t_{\rm phd}$ is in yr and $M_{\rm CO_{\rm obs}}$ in M$_{\oplus}$. The CO+CO$_2$ ice mass fraction in Fomalhaut is therefore in the range 4.6-76\% (taking the observed belt parameters quoted in Appendix \ref{sect:appb}). 

\subsection{Comparison with other CO-bearing systems: an overall similarity to Solar System comets}
\label{sect:comparison}

\begin{figure}
\vspace{-2mm}
 \hspace{-4mm}
  \includegraphics*[scale=0.365]{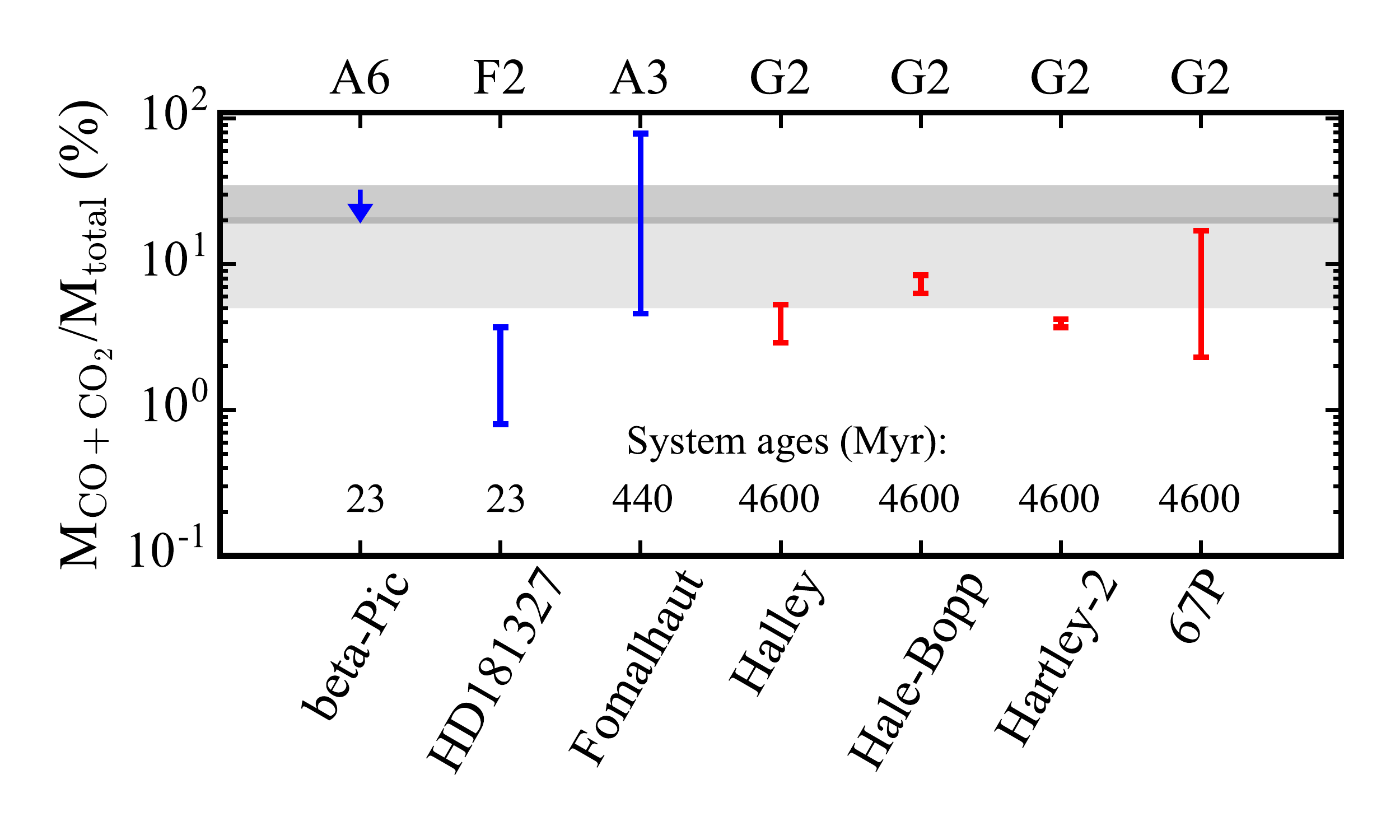}
\vspace{-7.0mm}
\caption{CO+CO$_2$ mass fraction (\%) in Solar System comets where both measurements are present \citep[red, assuming a dust to ice ratio of $\sim$4 as measured in comet 67P,][]{Rotundi2015} and in exocometary belts observed so far (blue, derived in the steady state framework). The tick labels on the top x axis indicate the spectral type of the host star. Compositions were derived from \citet{Matra2017} ($\beta$ Pic), \citet{Marino2016} (HD181327), this work (Fomalhaut), and \citet[][]{LeRoy2015} and references therein for Solar System comets. Shaded regions represent the ranges of mass fractions expected from direct inheritance of ISM compositions (assuming 100\% CO trapping within comets and no grain surface chemistry), in the cases where comets formed outside (darker grey) or inside (lighter grey) of the CO ice line within the protoplanetary disk.}
\label{fig:comcomp}
\end{figure}

Our compositional measurement in exocomets within the Fomalhaut ring adds to the the other two measurements of the CO+CO$_2$ exocometary ice mass fraction obtained through ALMA observations of second-generation CO gas, $\beta$ Pictoris \citep{Matra2017} and HD181327 \citep{Marino2016}. For consistency, we apply our updated, more accurate estimation of the ice mass fraction presented here also to these systems, using the same belt and stellar parameters as in the original works. We find that the CO+CO$_2$ ice mass fraction becomes $<32\%$ for $\beta$ Pictoris (using the 3$\sigma$ upper limit on the CO mass) and in the range 0.8-3.7\% for HD181327 (using the $\pm1\sigma$ range of the CO mass). These two systems both belong to the $\beta$ Pic moving group at an estimated age of 23$\pm$3 Myr \citep{Mamajek2014}, are much younger than Fomalhaut, but nonetheless present compositions that are within an order of magnitude of one another (Fig. \ref{fig:comcomp}). For example, the weak detection in Fomalhaut compared to $\beta$ Pic is easily explained by the fact that Fomalhaut 
is much less collisionally active. Indeed, the mass loss rate through the cascade is $\gtrsim20$ times higher for the $\beta$ Pictoris disk, and the higher amounts of CO also mean that some self-shielding takes place, increasing the photodissociation timescale by a factor $\sim$2.5 \citep{Matra2017}. Overall, for the same CO+CO$_2$ ice content, this indicates that the CO gas mass would be at least 50 times higher in $\beta$ Pic with respect to Fomalhaut, in agreement with the observations. On the other hand, we note that our new Fomalhaut measurement is at least marginally inconsistent with that in the HD181327 debris ring, potentially suggesting an intrinsic difference in composition, or gas release mechanism, between these two systems.

Despite the rather large error bars on the measurements obtained so far, Fig. \ref{fig:comcomp} also shows that the CO+CO$_2$ mass fraction in exocomets is consistent with that of Solar System comets.
A caveat to keep in mind in this comparison is that mass fractions in Solar System comets are derived from measurements of the relative abundance of CO and CO$_2$ compared to H$_2$O. On the other hand, their refractory to volatile mass ratio, required to derive the CO+CO$_2$ mass fraction, is only poorly (if at all) known, due to the difficulty in remote measurements of dust production rates. The only comet for which this is measured robustly in-situ is comet 67P/Churyumov-Gerasimenko (67P), where the dust to ice mass ratio is 4$\pm$2 in the coma \citep{Rotundi2015}, and consistent with (though not directly constrained by) density measurements of the interior of its nucleus \citep{Patzold2016}. We therefore assumed all Solar System comets to have an equal refractory to volatile mass ratio of 4, though note that this may be subject to scatter across comet families and individual objects.

Another underlying assumption is that we are using observation-based estimates of the gas production (outgassing) rates of CO and CO$_2$ (as well as H$_2$O for Solar System comets) as a probe for their relative solid abundance; in other words, we are assuming that CO+CO$_2$ ice and refractories are removed at rates that reflect their internal composition. In Solar System comets, linking outgassing rates in the comae and nuclei abundances depends not only on its distance from the star \citep[as clearly seen e.g. for Hale-Bopp,][]{Biver2002}, but also on its detailed thermal and structural properties \citep[][]{Marboeuf2012}. However, CO/H$_2$O and CO$_2$/H$_2$O outgassing rate ratios are generally measured near a comet's perihelion, where the process is dominated by water sublimation which carries along CO and CO$_2$ trapped within it. Then, for a comet with a mostly homogeneous composition, and dominated by clathrates \citep[as shown to hold for 67P,][]{ ,Luspay-Kuti2016}, outgassing ratios are expected to be a good representation of its ice abundance \citep[see e.g.][]{Marboeuf2014}.     

As with Solar System comets, our exocometary abundance derivation also assumes that CO and CO$_2$ ice, as well as refractories, are released at rates that reflect their composition; in the context of our steady state model, this is valid as long as two conditions apply:
\begin{enumerate}
\item No CO or CO$_2$ is removed as ice by way of the smallest blow-out grains in the cascade, which is of order $\sim$10 $\mu$m in the Fomalhaut belt (though it can vary depending on grain composition). 
In other words, these ices are only removed as gas. This applies if ices are completely lost further up the collisional cascade, i.e. if either the sublimation or photodesorption timescale is shorter than the collision timescale for grains larger than the blow-out size. Taking as an example a 20 $\mu$m-sized pure water ice grain, the results of \citet{Grigorieva2007} can be used to show that its photodesorption timescale is $\sim$13700 yrs. On the other hand, its collisional lifetime is $\sim4\times10^{5}$ yrs \citep{Wyatt2002}, meaning that water ice cannot survive on the surface of grains at the bottom of the collisional cascade. A similar argument applies for a pure CO$_2$ ice grain, whose photodesorption timescale will be very similar to that of water, due to a similar photodesorption yield of $\sim10^{-3}$ molecules photon$^{-1}$ \citep{Grigorieva2007, Oberg2009}.

The next question is whether CO gas or CO$_2$ ice can remain trapped within refractories all the way down to the smallest sizes in the cascade. Due to its low sublimation temperature ($\sim$20 K), CO is trapped already in gas form at the blackbody temperature of a dust grain within the Fomalhaut belt ($\sim$48K), and for small grains at the bottom of the cascade, it is likely to diffuse through the refractory layer. On the other hand, CO$_2$ has a higher sublimation temperature of $\sim$80 K \citep{Collings2004}. The hottest grains will be those near the blow-out limit ($\sim$10 $\mu$m), which will have a temperature close to that observed in the SED \citep[$\sim$74 K,][]{KennedyWyatt2014}. Further heating is likely to take place through collisions themselves \citep[e.g. with accelerated high-$\beta$ grains, as proposed by][]{Czechowski2007}, increasing the likelihood of CO$_2$ sublimation and subsequent release  through diffusion. In conclusion, while further detailed modelling is necessary, we deem it very unlikely for any CO to be retained within grains down to the smallest sizes in the cascade, but on the other hand cannot exclude that a fraction of CO$_2$ ice may instead be retained within a refractory mantle. The latter would cause an underestimate of the CO+CO$_2$ exocometary mass fractions presented here.



\item CO and CO$_2$ production is not dominated by resurfacing collisions, which preferentially occur for large bodies at the top of the cascade (Bonsor et al. in prep.). These less energetic collisions, which are not taken into account in our model, can expose trapped volatiles as well as fresh surface ice that can then be rapidly lost through sublimation and/or photodesorption. This would produce more gas mass than predicted through catastrophic-only planetesimal collisions, meaning that our model would be overestimating the cometary CO+CO$_2$ ice mass fraction. If resurfacing collisions are the main driver for gas release, we expect most of this release to happen early in the lifetime of the belt, since it requires the big planetesimals (holding most of the CO mass) to only have been resurfaced rather than destroyed.
 However, we note that these large bodies may also be large enough to retain the released gas through an atmosphere and/or could lose dust as well as gas through drag (as observed in Solar System comets). Overall, it remains to be established whether gas released through resurfacing collisions can dominate the released mass, though such a study is beyond the scope of this paper. 

\end{enumerate}    

\subsection{Possible origins for this similarity and comparison to a simple ISM inheritance model}
\label{sect:similorigin}

We find that CO+CO$_2$ mass fractions in exocomets are similar to each other and are of the same order (within about an order of magnitude) as observed for Solar System comets \citep[][and references therein]{LeRoy2015}. As shown in Fig. \ref{fig:comcomp}, this similarity appears to apply around host stars of a range of ages and spectral types. In terms of distance to the host star, exocometary gas is observed at 50-220 AU around a 8.7 L$_{\odot}$ star \citep[$\beta$ Pic,][]{Matra2017}, $\sim$81 AU around a 3.3 L$_{\odot}$ star \citep[HD181327,][]{Marino2016}, and $\sim$135 AU around a 16.6 L$_{\odot}$ star \citep[this work, stellar luminosity from SED fit in][]{KennedyWyatt2014}. Assuming blackbody-like bodies, the equilibrium temperature in these belts would be equivalent to distances of 17-75, 45 and 33 AU from the Sun in the Solar System, meaning that the temperature of these exocomets should be of the same order as observed for the Solar System's comet reservoir in the Kuiper belt \citep[30-50 AU, e.g.][]{Stern1997}. Then, similar compositions and temperature environments for comets may indicate similar comet formation conditions in younger protoplanetary disks including the Solar nebula. 

Another aspect to consider is whether these cometary fractions of mass locked in CO and CO$_2$ are globally representative of the chemical heritage from the interstellar medium \citep[ISM, see][for an extensive discussion of such inheritance]{Pontoppidan2014}. In the ISM, we know that the [CO/H$_2$] abundance ratio in the gas is of order $10^{-4}$, where H$_2$ dominates the gas mass, and the gas/dust ratio is of order $\sim$100. This yields a M$_{\rm CO gas, ISM}$/M$_{\rm total, solids, ISM}\sim14\%$. On the other hand, the CO abundance in ISM ices is in the range [CO/H$_2$O]$_{\rm ice, ISM}\sim$9-36\% compared to H$_2$O \citep[][and references therein]{Mumma2011}.
The CO$_2$ content is dominated by its ice phase \citep[e.g.][]{vanDishoeck1996}, with a [CO$_2$/H$_2$O]$_{\rm ice, ISM}$ abundance of $\sim$15-44\% \citep[again, see][and references therein]{Mumma2011}. This means that ISM material that is accreted in the outer regions of the protoplanetary disk will contain CO+CO$_2$ already in the ice form, with M$_{\rm CO+CO_2 ice, ISM}$/M$_{\rm H_2O ice, ISM}\sim$0.51-1.64, and CO in the gas form, with M$_{\rm CO gas, ISM}$/M$_{\rm total, solids, ISM}\sim0.14$.

Given their present location, we assume that cometary belts formed in protoplanetary disks outside the H$_2$O and CO$_2$ ice lines. This means that without significant vertical and outward radial mixing H$_2$O and CO$_2$ remained locked on the grains with ISM abundances, producing cometary CO$_2$/H$_2$O abundances representative of the ISM. For the CO content, however, we consider two possible scenarios, where comets formed either outside or inside the CO ice line. Outside of the CO ice line, the freeze-out of CO from ISM gas will enhance the ice-phase CO+CO$_2$ abundance compared to the ISM value. This simple scenario would yield a total CO+CO$_2$ mass fraction in the comets given by
\begin{equation}
\left(\frac{\rm{M_{CO+CO_2, ice}}}{\rm{M_{total, solids}}}\right)_{\rm comet}=\frac{\left(\rm{M_{CO+CO_2, ice}}\right)_{\rm ISM}+\left(\rm{M_{CO, gas}}\right)_{\rm ISM}}{\left(\rm{M_{total, solids}}\right)_{\rm comet}}
\end{equation}
where
\begin{equation}
\frac{\left(\rm{M_{CO+CO_2, ice}}\right)_{\rm ISM}}{\left(\rm{M_{total, solids}}\right)_{\rm comet}}=\frac{1}{1+\left(\frac{\rm{M_{dust}}}{\rm{M_{ice}}}\right)_{\rm comet} }  \frac{1}{1+\left( \frac{\rm{M_{CO+CO_2, ice}}}{\rm{M_{H_2O, ice}}}\right)^{-1}_{\rm ISM}}
\end{equation}
and 
\begin{equation}
\frac{\left(\rm{M_{CO, gas}}\right)_{\rm ISM}}{\left(\rm{M_{total, solids}}\right)_{\rm comet}}=\frac{1}{1+\left(\frac{\rm{M_{CO, gas}}}{\rm{M_{total, solids}}}\right)^{-1}_{\rm ISM} }
\end{equation}
Given CO gas and CO+CO$_2$ ice abundances in the ISM, assuming a range of dust to ice ratios between the $\pm1\sigma$ values measured in comet 67P, we obtain an expected CO+CO$_2$ cometary mass fraction of $\sim$17-33\% (darker shaded region in Fig. \ref{fig:comcomp}).
Growth of grains to comet-sized bodies allows CO ice originally on grain mantles to become trapped in other ices and refractories; this ensures that CO (and thus the CO+CO$_2$ mass fraction derived) can be retained once the protoplanetary disk is dispersed and the CO ice line moves further out than the cometary belt location.

In the second formation scenario, where the cometary belt formed inside the CO ice line in the protoplanetary disk, no extra CO from ISM-like gas would be incorporated in the ice phase; this would imply a CO+CO$_2$ cometary mass fraction of $\sim$5-21\% (lighter shaded region in Fig. \ref{fig:comcomp}). As well as assuming that all of the CO can remain trapped within other ices or refractories inside of its ice line, this simplified evolutionary model ignores any grain surface chemistry, which is likely ongoing through the ISM, protoplanetary and cometary phases of evolution. We expect that such chemistry will act to deplete the CO and CO$_2$ ice abundances, creating not only more complex volatiles (which are not dominant in cometary ice), but also organic refractories. Strictly speaking, we should therefore consider these ranges as upper limits to the CO+CO$_2$ cometary mass fractions derived in this ISM inheritance model.

While the large error bars in the Solar System and exo- cometary measurements do not allow us to draw significant conclusions with regards to enhancement or depletion compared to ISM-inherited abundances, we  show that such a comparison should be possible with increasingly accurate observations. For example, measuring the depletion of CO+CO$_2$ abundance with respect to ISM-inherited values would allow us to estimate the amount of CO and/or CO$_2$ that has been lost either due to sublimation during or immediately after dispersal of the protoplanetary disk (due to consequent outward movement of the CO ice line location) or due to grain surface chemistry and production of more complex organics.  As well as achieving more accurate measurements over a larger sample of exo- and Solar System comets, pinpointing the formation location of cometary belts within the protoplanetary disk with respect to the CO ice line will be crucial in allowing us to distinguish between the two possible ranges of ISM-inherited abundances discussed above.



\subsection{Pericentre/apocentre glow of CO released from a steady state collisional cascade}
\label{sect:COperiglow}

In Sect. \ref{sect:morph} we presented tentative evidence of an enhancement in CO emission near the planetesimal belt's pericentre. Here, we examine its possible physical origin in the framework of our steady state model described in Sect. \ref{sect:stst}. Once again, we assume that the CO gas produced is a fraction of the solid mass lost as part of a steady state collisional cascade, giving us access to the CO+CO$_2$ ice abundance. The total solid mass loss rate $\dot{M}(D)$ is calculated by multiplying the mass $M_{\rm solid}(D)$ in any given size bin with the collision rate $R_{\rm col}(D)$ of bodies in that size bin.

\begin{figure}
 \hspace{-5mm}
  \includegraphics*[scale=0.31]{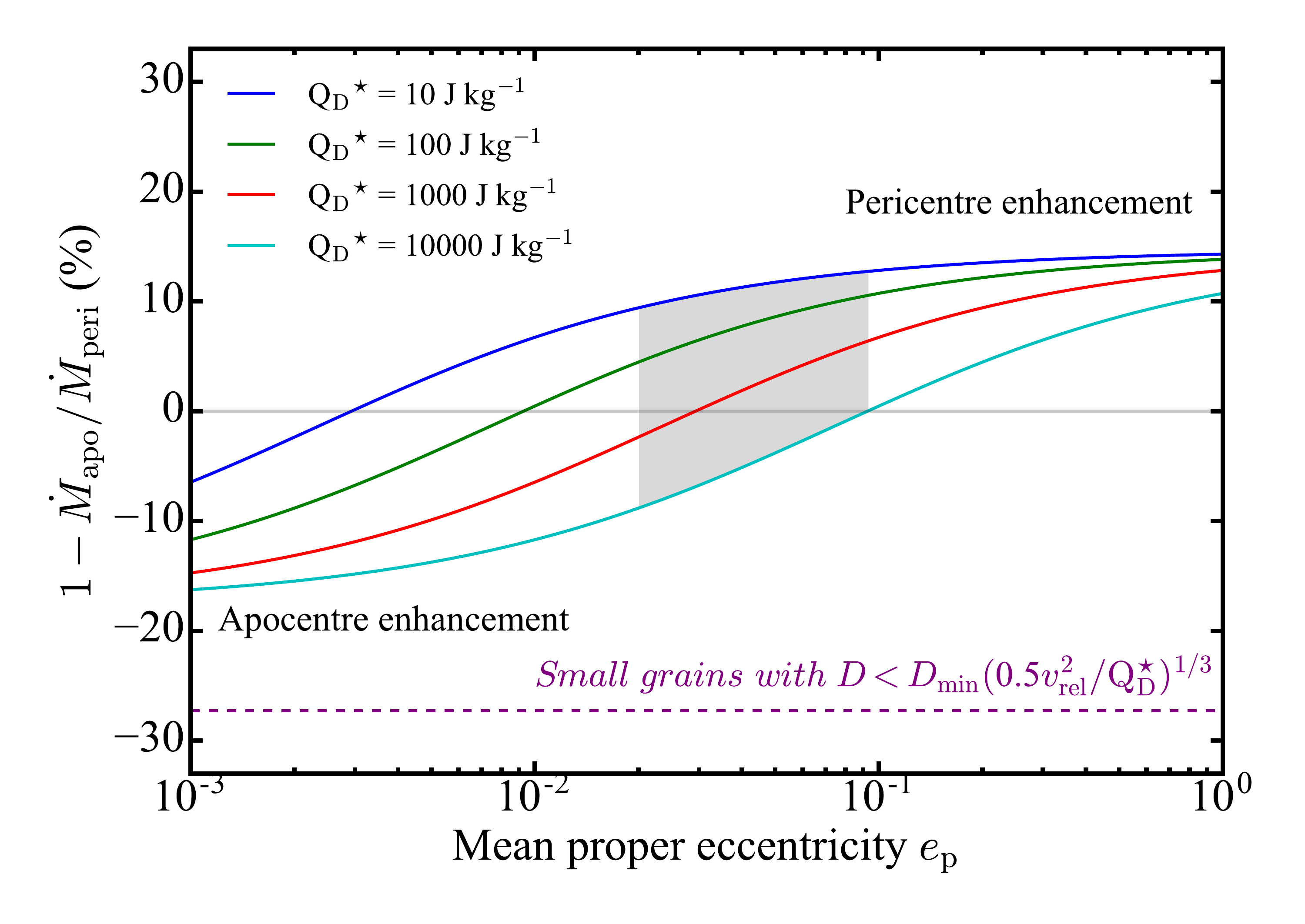}
\vspace{-8.0mm}
\caption{Solid mass loss rate (and hence CO mass) enhancement predicted at pericentre with respect to apocentre in the Fomalhaut ring, for CO released by small grains of sizes $D<D_{\rm min}($0.5$v_{\rm rel}^2/{\rm Q}_{\rm D}^{\star})^{1/3}$ (purple dashed line), or by larger grains of sizes $D>D_{\rm min}($0.5$v_{\rm rel}^2/{\rm Q}_{\rm D}^{\star})^{1/3}$ (coloured solid lines). In the larger size regime, given the observationally constrained $M_{\star}, q, a, e_{\rm frc}$ (see text), the enhancement is only a function of the unknown proper eccentricity $e_{\rm p}$ of the planetesimals and of their threshold specific strength needed for catastrophic collisions ${\rm Q}_{\rm D}^{\star}$. 
The shaded region represents a wide range of ${\rm Q}_{\rm D}^{\star}$ as explored by \citep{Wyatt2002} and the $\pm1\sigma$ interval of $e_{\rm p}$ obtained from fitting the ALMA dust image \citep{Macgregor2017}.}
\label{fig:periapoenh}
\end{figure}

In Appendix \ref{sect:appa}, we quantify how this mass loss rate within an eccentric cometary belt depends on true anomaly $f$, for two possible regimes of grain sizes releasing CO.
For the small grains, under the condition $D<D_{\rm min}($0.5$v_{\rm rel}^2/{\rm Q}_{\rm D}^{\star})^{1/3}$, we find that the collision rate is independent of true anomaly; this in turn means that the mass loss rate should be enhanced at apocentre due to the expected enhancement in solid mass at this location (due to particles spending more time there). For larger grains ($D>D_{\rm min}($0.5$v_{\rm rel}^2/{\rm Q}_{\rm D}^{\star})^{1/3}$) the azimuthal 
dependence of the mass loss rate is set by 6 parameters, namely the slope of the size distribution in the collisional cascade $q$, the mean proper eccentricity of material orbiting within the belt e$_{\rm p}$, the stellar mass M$_{\star}$, the belt semimajor axis a, the forced eccentricity of the belt e$_{\rm frc}$, and the specific incident energy required for a catastrophic collision, ${\rm Q}_{\rm D}^{\star}$.
For the case of the Fomalhaut belt, we have observational constraints on $q\sim1.83$ \citep[][,where we assume it to be independent of true anomaly]{Ricci2012}, $M_{\star}\sim1.92$ M$_{\odot}$ \citep{Mamajek2012}, $a\sim136.3$ AU and $e_{\rm frc}\sim0.12$ \citep{Macgregor2017}, meaning that our only free parameters are the mean proper eccentricity $e_{\rm p}$ and the specific strength ${\rm Q}_{\rm D}^{\star}$ of the planetesimals. 
We can then quantify the fractional difference in mass loss rate at pericentre with respect to apocentre ($1-\dot{M}(f=180^{\circ})/\dot{M}(f=0^{\circ})$) using Eq. \ref{eq:mlossvkep} and \ref{eq:vkep} in Appendix \ref{sect:appa}, shown for a wide range of $e_{\rm p}$ and ${\rm Q}_{\rm D}^{\star}$ values in Fig. \ref{fig:periapoenh}. 

As expected, for low proper eccentricities and high planetesimal strengths the effect of a higher mass at apocentre dominates over the effect of easier collisional disruption at pericentre. Vice versa, for high proper eccentricities and low planetesimal strengths, we see a pericentre enhancement in the mass loss rate due to the effect of easier collisional disruption at pericentre dominating over the effect of having a higher mass at apocentre. The upper and lower horizontal asymptotes of the curves correspond to the limits $e_{\rm p}\rightarrow0$ and $e_{\rm p}\rightarrow\infty$ as calculated by setting $e_{\rm frc}=0.12$ in Eq. \ref{eq:ezerolim} and \ref{eq:einflim} (Appendix \ref{sect:appa}), and are independent of both $e_{\rm p}$ and ${\rm Q}_{\rm D}^{\star}$. This implies a $\sim$15\% maximum pericentre to apocentre enhancement in the mass loss rate of the Fomalhaut belt.
For the $\pm1\sigma$ range of proper eccentricities derived from continuum imaging \citep{Macgregor2017}, and ${\rm Q}_{\rm D}^{\star}$ values in the range 10-10000 J kg$^{-1}$, expected for weak ice grains of a wide range of sizes within the collisional cascade \citet{Wyatt2002} \citep[assuming the results of][]{BenzAsphaug1999}, we predict the fractional difference in mass loss rate between pericentre and apocentre in the range -9 to 12\%.
Then, given that the CO photodissociation rate is independent of true anomaly, and assuming that the CO+CO$_2$ ice fraction also is, the steady state mass loss rate enhancement at pericentre directly translates into a CO mass enhancement.

Whether a CO mass enhancement translates into a CO flux enhancement at pericentre/apocentre, or in other words \textit{CO pericenter/apocentre glow}, also depends on the excitation of the CO molecule (see Sect. \ref{sect:B7check}), and particularly on the radial dependence of the gas kinetic temperature and electron density. This is because for example in Fomalhaut, CO emitted at the ring's pericentre will be closer to the star than CO at apocentre, by a factor $(1+e)/(1-e)\sim1.27$. Making the simple assumption of a $\beta$ Pic-like environment for CO excitation, i.e. with an electron density varying with radius as $\sim 300$ $(R/100 {\rm AU})^{-1}$ cm$^{-3}$ \citep{Matra2017}, the expected flux at pericentre would be 2.5-6.1\% larger than that at apocenter, making the CO flux enhancement even larger than expected from a mass enhancement only. Therefore, we expect excitation effects to favor a CO flux enhancement at pericentre. We cautiously note that such contribution from CO excitation will depend on the gaseous environment of the Fomalhaut belt; this may significantly differ from that of $\beta$ Pic, though it may be characterised in future using optically thin line ratio observations.

Finally, we need to take into account projection effects due to the viewing geometry; in particular, the observed pericenter to apocenter ratio will depend on the combined effect of the ring's vertical thickness, its inclination to the line of sight and the on-sky angular distance between the ring's pericentre and the nearest ansa. This is because the azimuthal intensity distribution for a significantly inclined ring with a non-negligible vertical thickness should present two enhancements at the location of the two ansae \citep[see][for the dependence of the azimuthal intensity distribution on a ring's scale height]{Marino2016}. To fully take this effect into account, and demonstrate its impact on the measured pericentre or apocentre glow, we produce a simple model of CO in the Fomalhaut ring and produce a sky-projected model image using the \texttt{RADMC-3D}\footnote{\url{http://www.ita.uni- heidelberg.de/dullemond/software/radmc-3d}} radiative transfer code. In doing this, we first construct an eccentric CO ring model with radial mass distribution, inclination to the line of sight, position angle, forced eccentricity and argument of pericentre equal to those of the dust ring. Then, we introduce the dependence of the CO surface density on the true anomaly as described above and derived in Appendix \ref{sect:appa}. We assume the maximum possible pericentre vs apocentre CO mass enhancement for the best-fit $e_{\rm p}\sim0.06$ from the continuum observations, corresponding to $\sim$12\% (for a ${\rm Q}_{\rm D}^{\star}$ of 10 J kg$^{-1}$). Furthermore, we make the simple assumption that the gas kinetic temperature is equal to the blackbody temperature of the planetesimals, and that the electron density follows the same radial dependence as found in $\beta$ Pictoris. 

\begin{figure}
 \hspace{-9mm}
  \includegraphics*[scale=0.4]{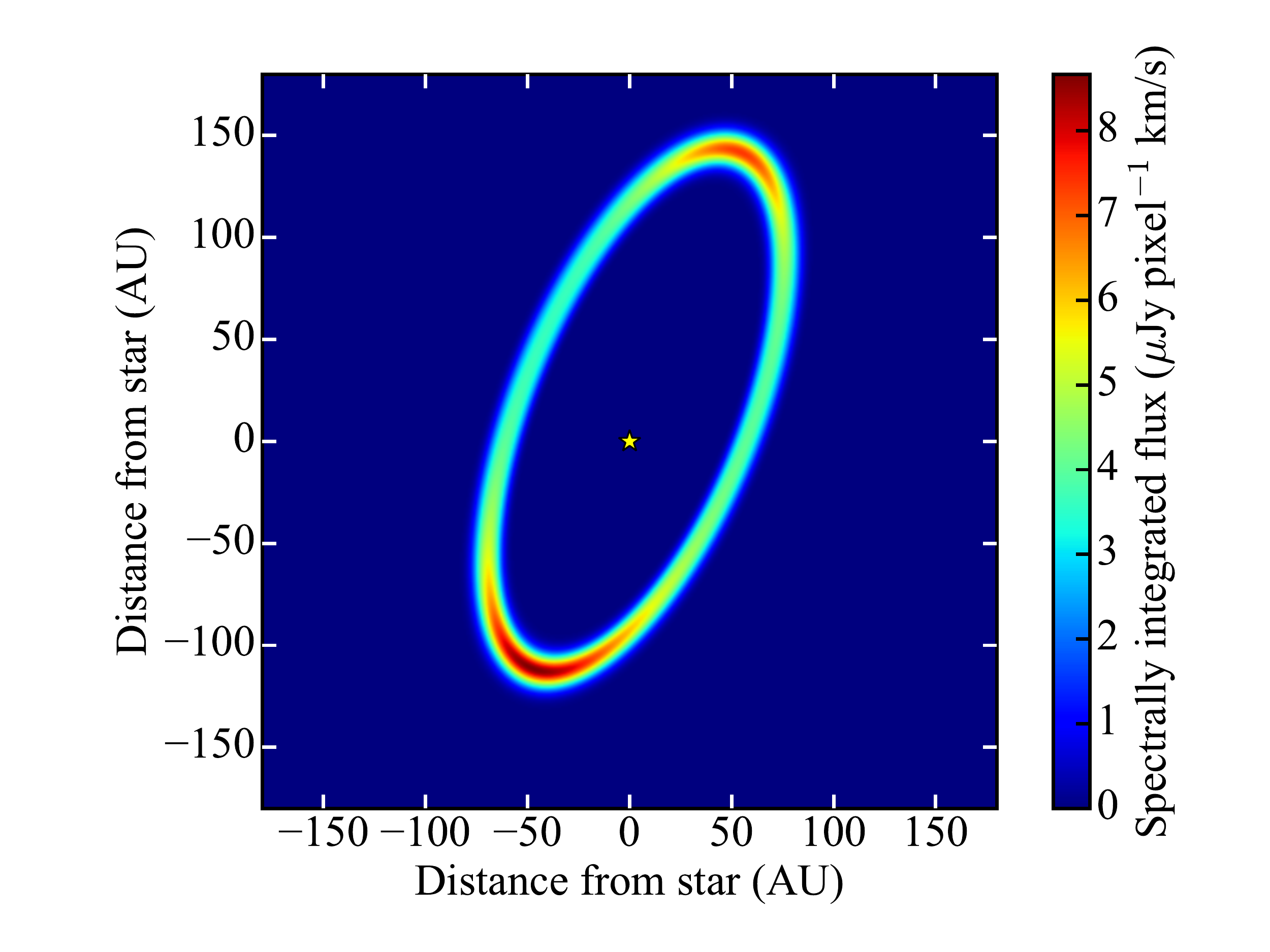}
\vspace{-6mm}
\caption{Model image for the maximum CO J=2-1 flux enhancement expected at pericentre with respect to apocentre through steady state CO production in the Fomalhaut ring. The model predicts a flux density enhancement of $\sim14.5$\% at the pericentre ($\omega=22\fdg5$) with respect to the apocentre location. However, this is for an infinitesimally short CO lifetime, much shorter than an orbital timescale; for the finite expected CO lifetime of 120 years, the peak is shifted by $\sim$19$^{\circ}$ in the direction of motion, which we here assumed to be clockwise. This shifts the peak to a true anomaly of $f\sim3\fdg5$, i.e. very close to the SE ansa. This model example assumes the best-fit orbital elements and ring width from the dust continuum fit, $\beta$ Pictoris-like electron densities, a gas temperature equal to the blackbody temperature of $\sim50$K at the ring's distance to the star, a planetesimal strength ${\rm Q}_{\rm D}^{\star}=10$ and a radially constant vertical aspect ratio of 0.06.}
\label{fig:modelpred}
\end{figure}

In our model (see spectrally integrated image in Fig. \ref{fig:modelpred}), we assume a radially constant CO vertical aspect ratio equal to the average proper eccentricity of the planetesimals ($h\sim0.06$), but we also consider extreme cases of a very small aspect ratio ($h\sim0.0001$) and a larger one \citep[$h\sim0.14$, corresponding to the 3$\sigma$ upper limit found for the HD181327 ring,][]{Marino2016}. We find that in the infinitesimally vertically thin limit we recover the pericentre enhancement with respect to apocentre expected from our CO steady state model from both mass (12\%) and excitation ($\sim$2.5\%) effects. In the vertically thick cases the value of this ratio is slightly reduced due to the additional `background' effect of the ansae enhancement. Their absolute flux difference, however, remains unchanged due the projection effect being axisymmetric with respect to the ring's geometric centre. 

To conclude, the upper limit pericentre vs apocentre enhancement predicted by our model (14.5\%) is marginally consistent with our measurement from the current dataset ($88\pm25$ \%, see Sect. \ref{sect:morph}) at the 2.9$\sigma$ level. If confirmed at high significance, this discrepancy would indicate the inability of our steady state model to explain this asymmetry. In turn, this may favour a stochastic event such as the destruction of a large icy body near the belt's pericentre (as discussed in Sect. \ref{sect:stoch}). Similarly, a recent impact was previously invoked to explain the observed CO asymmetry in the $\beta$ Pictoris system \citep{Dent2014}, though this was recently ruled out through higher resolution observations \citep{Matra2017}.

Either way, conclusive detection of this enhancement would allow us to determine the direction of orbital motion of the belt, and in turn whether its East or West side is nearest to Earth. This is because the finite lifetime of CO ($\sim$120 years in such optically thin environment) would cause either a tail in the direction of motion \citep[in case of a stochastic event, see e.g.][]{Dent2014}, or a $\sim$19$^{\circ}$ offset of the peak location (with respect to pericentre/apocentre, also in the direction of motion) due to CO pericentre/apocentre glow. 
Therefore future, deeper ALMA observations of pericentre and apocentre are warranted to confirm this tentative evidence for asymmetry.
\\

\section{Conclusions}
\label{sect:concl}

In summary, this work presented observations of CO J=2-1 230 GHz emission in the Fomalhaut ring. Through spectro-spatial filtering of the ALMA data cube, we detected line emission with an integrated line flux of $68\pm16$ mJy km/s at a radial velocity consistent with that of the star. We report the following findings:
\begin{enumerate}

\item The spectro-spatial filtering method shows that the ring's sky-projected rotation axis matches that of the star \citep{LeBouquin2009}, with the material at the SE ansa moving towards us. This however remains insufficient to determine the sense of rotation and in turn whether the NE side of the ring \citep[observed to be brighter in HST scattered light imaging,][]{Kalas2005} is in front or behind the sky plane.

\item The radial location of the emission is consistent with that of the millimetre dust ring \citep[as presented in][]{Macgregor2017}, indicating that both the observed CO and dust originate from the cometary belt at a distance of $\sim136$ AU from the star.

\item CO J=2-1 emission is optically thin and originates from a total CO gas mass of $0.65-42 \times10^{-7}$ M$_{\oplus}$. This is consistent with the previous ALMA non-detection of the J=3-2 transition \citep{Matra2015}, and is in line with the CO excitation conditions observed in the $\beta$ Pictoris disk.

\item At an age of 440 Myr, Fomalhaut hosts the oldest debris belt where gas emission co-located with dust emission has been detected to date. The amount of CO and the potential high amounts of H$_2$ (more typical of primordial protoplanetary disks) are insufficient to shield CO and allow it to survive over the system's lifetime. This implies that the observed CO is of secondary origin and originates from exocometary ices within the belt.

\item We evaluate the possibility of CO being produced from either stochastic destruction of a large icy body or steady state release through the collisional cascade. We deem a stochastic collision possible but reasonably likely only for high total belt masses. On the other hand, in the steady state scenario \citep[as first described in][]{Matra2015}, we combine the mass loss rate from the collisional cascade (producing CO) with the known CO photodissociation rate (destroying CO) to infer a CO+CO$_2$ exocometary mass fraction of 4.6-76\%. This is consistent with the other two debris belts where gas has been confirmed to be of exocometary origin, $\beta$ Pictoris \citep{Matra2017} and HD181327 \citep{Marino2016}.

\item As well as being similar to one another, exocometary CO+CO$_2$ mass fractions are consistent with observations of Solar System comets, where this may be explained by similar blackbody temperatures and may indicate similar formation conditions in the original protoplanetary disk. We present a simple ISM inheritance model, showing that the CO+CO$_2$ mass fractions in exo- and Solar System comets are consistent with all of the CO+CO$_2$ having been directly inherited from the ISM's CO+CO$_2$ ice and CO gas content. Increasingly accurate cometary abundance measurements are needed to distinguish between comet formation scenarios and to estimate the amount of CO+CO$_2$ that was lost through grain surface chemistry forming more complex organics and/or through gas release during the main sequence phase of evolution, as observed here.

\item We report tentative evidence that most of the detected CO emission (49$\pm$27)\% originates near the ring's pericentre location derived by ALMA and HST high-resolution dust imaging. This may be due to a recent impact event that took place near pericentre, or to CO pericentre glow caused by the combined effect of A) a steady state mass loss rate enhancement at pericentre for an eccentric ring (for sufficiently high values of proper eccentricities $e_{\rm p}$ and/or sufficiently low planetesimal strengths ${\rm Q}_{\rm D}^{\star}$); and B) molecular excitation effects due to the pericentre being closer to the central star.

\item We presented a model of CO pericentre or apocentre glow expected for exocometary CO released in eccentric belts. For a well characterised dust belt such as Fomalhaut, the expected CO mass ratio at apocentre with respect to pericentre is only a function of the planetesimal strength ${\rm Q}_{\rm D}^{\star}$ and the mean proper eccentricity $e_{\rm p}$. The model presented is general and indicates that we should expect asymmetric exocometary emission in eccentric gas-bearing debris disks, a prediction that may be tested by future ALMA observations. The highest possible CO flux enhancement at pericentre vs apocentre predicted in the Fomalhaut belt ($\sim14.5$\%) is at the limit of being marginally consistent with our observations; if confirmed, a pericentre enhancement much higher than this prediction would rule out a steady state scenario, proving instead that the observed CO must have originated from a recent impact between very large comets.

\end{enumerate}




\acknowledgments
The authors are grateful to the referee for his/her thorough review of the paper, leading to significant and valuable improvements. In addition, we would like to thank S. Marino and A. Bonsor for useful discussions on the filtering method and the cometary release mechanism.  LM acknowledges support by STFC through a graduate studentship and, together with MCW and AS, by the European Union through ERC grant number 279973. M.A.M acknowledges support from a National Science Foundation Graduate Research Fellowship (DGE1144152) and from NRAO Student Observing Support. PK, JRG and GD thank NSF AST-1518332, NASA NNX15AC89G and NNX15AD95G. This work benefited from NASA's Nexus for Exoplanet System Science (NExSS) research coordination network sponsored by NASA's Science Mission Directorate. GMK is supported by the Royal Society as a Royal Society University Research Fellow. AMH gratefully acknowledges support from NSF grant AST-1412647. MP acknowledges support from NASA grants NNX15AK23G and NNX15AM35G. AS is partially supported by funding from the Center for Exoplanets and Habitable Worlds. The Center for Exoplanets and Habitable Worlds is supported by the Pennsylvania State University, the Eberly College of Science, and the Pennsylvania Space Grant Consortium. This paper makes use of the following ALMA data: ADS/JAO.ALMA\#2015.1.00966.S. ALMA is a partnership of ESO (representing its member states), NSF (USA) and NINS (Japan), together with NRC (Canada), NSC and ASIAA (Taiwan), and KASI (Republic of Korea), in cooperation with the Republic of Chile. The Joint ALMA Observatory is operated by ESO, AUI/NRAO and NAOJ.




Facilities: \facility{ALMA}.

\bibliographystyle{apj}
\bibliography{lib}

\appendix

\section{A. Dependence of mass loss rate on true anomaly in a steady state collisional cascade}
\label{sect:appa}

Here, we aim to derive the dependence of the solid mass loss rate $\dot{M}(D,f)=M(D,f)R_{\rm col}(D,f)$ on the true anomaly $f$ in an element of length along the ring $ds$ and in a size bin between $D$ and $D+dD$.
Following \citet{Wyatt2002}, and explicitly marking the dependence of parameters on the true anomaly $f$, the collision rate of planetesimals of sizes in the range $D_{\rm im}$ to $D_{\rm im}$+d$D_{\rm im}$ on a planetesimal of size $D$ is
\begin{equation}
R_{\rm col}(D, D_{\rm im}, f)=F(D, D_{\rm im})\sigma(f) v_{\rm rel}(f)
\end{equation}
where $\sigma(f)$ is the cross-sectional area per unit volume of planetesimals of all sizes, $v_{\rm rel}(f)$ is the relative velocity of the impactor and the target, and $F(D, D_{\rm im})$ is the collisional cross-section of impactors of size $D_{\rm im}$ on the target of size $D$.
Neglecting gravitational focusing, which in Fomalhaut becomes important only for bodies that are too large to participate in the collisional cascade \citep{Wyatt2002}, the latter is expressed as
\begin{equation}
F(D, D_{\rm im})=\bar{\sigma}(D_{\rm im})\left(1+\frac{D}{D_{\rm im}}\right)^2
\end{equation}
where $\bar{\sigma}(D_{\rm im})=\sigma(D_{\rm im})/\int_{D_{\rm min}}^{D_{\rm max}}\sigma(D)dD$ is the normalised cross section of the impactor, with $D_{\rm min}$ and $D_{\rm max}$ being the minimum and maximum size of solids participating in the collisional cascade.

The catastrophic collision rate for a planetesimal of size D from impactors of all sizes is then
\begin{equation}
 \label{eq:collrate}
R_{\rm col}(D,f)=\sigma(f) v_{\rm rel}(f)\int_{{\rm max}\{D_{\rm cc} (f,D) ,D_{\rm min}\}}^{D_{\rm max}}F(D, D_{\rm im}){\rm d}D_{\rm im}
\end{equation}
where $D_{\rm cc}(f,D)$ is the minimum impactor size for a collision to be catastrophic, i.e.
\begin{equation}
\label{eq:dcc}
D_{\rm cc}(f,D)=D\left(2{\rm Q}_{\rm D}^{\star}/ v_{\rm rel}^2(f)\right)^{1/3}
\end{equation}
where ${\rm Q}_{\rm D}^{\star}$ is the specific incident energy required for a catastrophic collision, i.e. one where the largest collisional fragment has half the mass of the original target planetesimal.

We here assume that the relative velocity of planetesimals can be expressed as $v_{\rm rel} (f)=v_{\rm Kep} (f) \sqrt{1.25e_{\rm p}^2+I^2}$ \citep[e.g.][]{Lissauer1993}, and that the mean planetesimal inclinations $I\approx e_{\rm p}$, where $e_{\rm p}$ is the mean proper eccentricity of planetesimals in the belt. This differs from the forced eccentricity $e_{\rm frc}$ in that the forced component can be seen as that imposed on all particle orbits through secular interaction with an unseen perturber, whereas the proper component can be interpreted as the `intrinsic' eccentricity of the particle, and defines the width of the torus formed by particles orbiting the star with the same semimajor axis \citep[see e.g. Fig. 2 in][]{Wyatt1999}. Therefore, while the mean proper eccentricity $e_{\rm p}$ determines the extent of orbit crossing and changes the relative velocity of collisions equally at any true anomaly, it is the forced eccentricity $e_{\rm frc}$ that introduces a dependence on true anomaly on the Keplerian velocity, and hence on the relative velocity of collisions (see Eq. \ref{eq:vkep}).

Following \citet{Pan2016}, the linear number density, and hence the cross-sectional area density $\sigma(f)$ and the mass $M(D,f)$ in an element of length ds, depend on the inverse of the Keplerian velocity $\sigma(f) \propto 1/v_{\rm Kep} (f)$. This means that the product $\sigma(f) v_{\rm rel}(f)$ in Eq. \ref{eq:collrate} is independent of the Keplerian velocity $v_{\rm Kep}$ and hence of the true anomaly $f$. However, the collision rate $R_{\rm col}(D)$ remains dependent on $f$ through the minimum size $D_{\rm cc} (f,D)$ of impactors causing a catastrophic collision, which depends on the relative collision velocities $v_{\rm rel} (f)$.

As shown in \citet{Wyatt1999}, assuming that the steady state size distribution follows the expression $n(D)\propto D^{2-3q}$ from $D_{\rm min}$ to $D_{\rm max}$ \citep{Dohnanyi1969}, the integral in Eq. \ref{eq:collrate} can be solved to obtain
\begin{equation}
\label{eq:bigFinteg}
\int_{{\rm max}\{D_{\rm cc} (f,D) ,D_{\rm min}\}}^{D_{\rm max}}F(D, D_{\rm im}) {\rm d}D_{\rm im}=\left(\frac{X(f)D}{D_{\rm min}}\right)^{5-3q}\left[1+\frac{6q-10}{(3q-4)X(f)}+\frac{3q-5}{(3q-3)X(f)^2}\right], 
\end{equation}
where $X(f)=D_{\rm cc} (f,D)/D=\left(2{\rm Q}_{\rm D}^{\star}/ v_{\rm rel}^2(e_{\rm p},f)\right)^{1/3}$ for $D_{\rm cc} (f,D)>D_{\rm min}$ and $X(f)=D_{\rm min}/D$ for $D_{\rm cc} (f,D)<D_{\rm min}$.
One can immediately notice that for sizes where all impactors above the minimum size always cause a catastrophic collision ($D_{\rm cc} (f,D)<D_{\rm min}$), the integral becomes simply the expression in the square brackets above, and its dependence on true anomaly is lost. This has implications for the smallest particles in the size distribution, which we will come back to later.
For (larger) sizes where $D_{\rm cc} (f,D)>D_{\rm min}$, assuming that the smallest and largest sizes $D_{\rm min}$ and $D_{\rm max}$ as well as the slope of the collisional cascade parametrised by $q$ are independent of true anomaly, the azimuthal dependence of the collision rate can be expressed through $v_{\rm rel}$ as
\begin{equation}
R_{\rm col}(D,f) \propto v_{\rm rel}(f)^{2q-\frac{10}{3}}\left[1+\frac{6q-10}{\left(2{\rm Q}_{\rm D}^{\star}\right)^{\frac{1}{3}}(3q-4)}v_{\rm rel}(f)^{\frac{2}{3}}+\frac{3q-5}{\left(2{\rm Q}_{\rm D}^{\star}\right)^{\frac{2}{3}}(3q-3)}v_{\rm rel}(f)^{\frac{4}{3}}\right].
\label{eq:rcolvrelq}
\end{equation} 
Since the dependence on true anomaly enters the expression through the Keplerian velocity $v_{\rm Kep}$, we can rewrite the above the expression in terms of $v_{\rm Kep}(f)$. Then, we assume a \citet{Dohnanyi1969} size distribution ($q=11/6$), and include the $v_{\rm Kep}^{-1}$ dependence of the mass $M(D,f)$ in the belt to obtain the dependence of the mass loss rate on $v_{\rm Kep}$, namely
\begin{equation}
\label{eq:mlossvkep}
\dot{M}(D,f)\propto \left(\sqrt{2.25}e_{\rm p}\right)^{\frac{1}{3}}v_{\rm Kep}(f)^{-\frac{2}{3}}\left[1+\frac{2}{3}\left(2{\rm Q}_{\rm D}^{\star}\right)^{-\frac{1}{3}}\left(\sqrt{2.25}e_{\rm p}v_{\rm Kep}(f)\right)^{\frac{2}{3}}+\frac{1}{5}\left(2{\rm Q}_{\rm D}^{\star}\right)^{-\frac{2}{3}}\left(\sqrt{2.25}e_{\rm p}v_{\rm Kep}(f)\right)^{\frac{4}{3}}\right],
\end{equation}
where the Keplerian velocity in the eccentric planetesimal belt can be expressed as \citep[following][]{Pan2016}
\begin{equation}
\label{eq:vkep}
v_{\rm Kep}(f)=\sqrt{\frac{GM_{\star}}{a}\frac{1+2e_{\rm frc}{\rm cos}f+e_{\rm frc}^2}{1-e_{\rm frc}^2}}, 
\end{equation}
where $G$ is the gravitational constant, $M_{\star}$ is the stellar mass, and $a$ is the semimajor axis of the belt.

Equations \ref{eq:mlossvkep} and \ref{eq:vkep} have implications for the predicted enhancement of pericentre with respect to apocentre (or vice versa), which we express as $1-\dot{M}(f=180^{\circ})/\dot{M}(f=0^{\circ})$ and show in Fig. \ref{fig:periapoenh}. 
In particular, it can be shown that
\begin{equation}
\label{eq:ezerolim}
e_{\rm p}\rightarrow0 \Rightarrow \frac{\dot{M}(f=0^{\circ})}{\dot{M}(f=180^{\circ})}\rightarrow\left(\frac{v_{\rm Kep}(f=0^{\circ})}{v_{\rm Kep}(f=180^{\circ})}\right)^{-\frac{2}{3}}=\left(\frac{1+e_{\rm frc}}{1-e_{\rm frc}}\right)^{-\frac{2}{3}}<1
\end{equation}
and
\begin{equation}
\label{eq:einflim}
e_{\rm p}\rightarrow\infty \Rightarrow \frac{\dot{M}(f=0^{\circ})}{\dot{M}(f=180^{\circ})}\rightarrow\left(\frac{v_{\rm Kep}(f=0^{\circ})}{v_{\rm Kep}(f=180^{\circ})}\right)^{+\frac{2}{3}}=\left(\frac{1+e_{\rm frc}}{1-e_{\rm frc}}\right)^{+\frac{2}{3}}>1,
\end{equation}
meaning that a pericentre mass loss rate enhancement is expected for high proper eccentricities (since the high relative collision velocities at pericentre dominate over the mass enhancement at apocentre), whereas an apocentre mass loss rate enhancement is expected for low proper eccentricities (since the mass enhancement at apocentre dominates over the increase in collision velocities at pericentre). We note that in the limit of small proper eccentricities ($e_{\rm p}\rightarrow0$), the minimum impactor size $D_{\rm cc}(f,D)$ to cause a catastrophic collision on a target $D$ is much larger than the target itself ($D_{\rm cc}(f,D)\gg D$). This would create a collisional cascade where only the smallest of two colliding bodies is destroyed, while neither is destroyed if their sizes are similar.

It is also interesting to note that for small particles that can be destroyed by impactors of all sizes down to the minimum size (i.e. for particles where $D_{\rm cc} (f,D)<D_{\rm min}$) the collision rate $R_{\rm col}$ is independent of true anomaly. This means that for sufficiently small particles the dependence of the mass loss rate on true anomaly is driven purely by the azimuthal distribution of mass, leading to $\dot{M}\propto v_{\rm Kep}(f)^{-1}$, and causing an apocentre enhancement even stronger than described by Eq. \ref{eq:ezerolim} in the $e_{\rm p}\rightarrow0$ limit.

\section{B. On the mass loss rate of the smallest bodies in the collisional cascade}
\label{sect:appb}

The mass loss rate $\dot{M}_{D_{\rm min}}$ at which the smallest grains in the collisional cascade are removed from the system can be estimated through the collisional mass loss rate of grains just above the minimum size $D_{\rm min}$ in the cascade. As we will show in this section, the latter is well constrained observationally, and can be calculated through their observed cross-section $\sigma_{D_{\rm min}}$, leading to their mass $M_{D_{\rm min}}$, multiplied by their collision rate $R_{\rm col}(D_{\rm min})$.

For a thin ring such as Fomalhaut, the total mass of the smallest grains (in $M_{\oplus}$) can be expressed as a function of their total cross-sectional area $\sigma_{D_{\rm min}}\sim\sigma_{\rm tot}$ (in AU$^2$, where we are assuming that these grains dominate the total observed cross-sectional area of the ring $\sigma_{\rm tot}$) through
\begin{equation}
M_{D_{\rm min}}=2.5\times10^{-9}\rho\sigma_{\rm tot}D_{\rm min},
\end{equation}
assuming grains of mass density $\rho$ in kg m$^{-3}$, with $D_{\rm min}$ in $\mu$m. The total cross-sectional area can be expressed observationally as $\sigma_{\rm tot}=4\pi R^2f$, where for Fomalhaut we take $R$ to be the semimajor axis of the ring in AU, and $f=L_{\rm IR}/L{\star}$ is the fractional luminosity of the ring. Then, we can assume $D_{\rm min}$ to be the blow-out size for blackbody grains around a star of a given luminosity and mass (in units of $L_{\odot}$ and $M_{\odot}$), leading to \citep[e.g.][]{Wyatt2008}
\begin{equation}
M_{D_{\rm min}}=6.7\times10^{-5}R^2fL_{\star}M_{\star}^{-1}.
\end{equation}
Around Fomalhaut, given a stellar mass of 1.92 $M_{\odot}$ \citep{Mamajek2012}, a best-fit belt semimajor axis of 143.0 AU \citep[from ALMA continuum observations,][]{Macgregor2017}, a stellar luminosity of 16.6 L$_{\odot}$ and fractional luminosity of $7.8\times10^{-5}$ \citep[from SED fitting,][]{KennedyWyatt2014}, we obtain a mass in small grains of $9.2\times10^{-4}$ M$_{\oplus}$, which is within a factor $2$ of other values quoted by \citet{Zuckerman2012} and \citet{Acke2012}.

The collision rate of grains of size $D_{\rm min}$ can also be calculated, using Eq. \ref{eq:collrate} and \ref{eq:bigFinteg}. These can be greatly simplified when considering particles impacting the smallest grains of the cascade, under the condition that $D_{\rm cc} (f,D_{\rm min})<D_{\rm min}$, or in other words collisions impacting the smallest grains (dominated by grains of the same size, since these dominate the belt's cross-sectional area) are always catastrophic. Given the definition of $D_{\rm cc}(f,D)$ (Eq. \ref{eq:dcc}), this is the case for $D_{\rm cc}(f,D_{\rm min})/D_{\rm min}=\left(2{\rm Q}_{D_{\rm min}}^{\star}/ v_{\rm rel}^2(e_{\rm p})\right)^{1/3}<1$, or ${\rm Q}_{D_{\rm min}}^{\star}<v_{\rm rel}^2(e_{\rm p})/2$. In the Fomalhaut ring, given a best-fit mean proper eccentricity of 0.06 \citep{Macgregor2017}, $v_{\rm rel}\sim0.31$ km/s leading to a condition ${\rm Q}_{D_{\rm min}}^{\star}\lesssim5\times10^{4}$ J/kg for the smallest grains in the cascade. Using the compilation of ${\rm Q}_{\rm D}^{\star}$ versus size values in \citet{KrijtKama2014}, we find that this is always the case for grains larger than the blow-out size (7.2 $\mu$m for compact SiO$_2$ blackbody grains, and 18.7 $\mu$m for pure water ice grains). This remains valid if the smallest grains in the cascade are larger than the blow-out limit \citep[as argued in][]{KrijtKama2014}.

Then, $X(f)=D_{\rm min}/D$ with $D=D_{\rm min}$ implies $X(f)=1$ in Eq. \ref{eq:bigFinteg}. Estimating the cross-section per unit volume of grains (in AU$^{-1}$) of all sizes as $\sigma=\sigma_{\rm tot}/V$ where $V$ is the volume of the ring in AU$^3$, this reduces the expression for the collision rate of the smallest grains (Eq. \ref{eq:collrate}, here in yr$^{-1}$) to
\begin{equation}
\label{eq:rcolsimpl1}
R_{\rm col}(D_{\rm min})=0.21\frac{\sigma_{\rm tot}v_{\rm rel}}{V}\left[1+\frac{6q-10}{(3q-4)}+\frac{3q-5}{(3q-3)}\right] \ \ {\rm for} \ \ {\rm Q}_{D_{\rm min}}^{\star}<v_{\rm rel}^2(e_{\rm p})/2,
\end{equation}
where $v_{\rm rel}$ is in km/s.
We immediately note that the collision rate of these grains is now independent of ${\rm Q}_{D_{\rm min}}^{\star}$, removing the considerable amount of uncertainty introduced if calculating this collision rate for the largest bodies in the cascade, as previously worked out in \citet{Matra2015} from the results of \citet{Wyatt2002}, as well as other works \citep{Kennedy2015,Marino2016, Matra2017, Kral2017}.
Expressing the cross-sectional area as a function of the belt's fractional luminosity, inserting the definition of $v_{\rm rel}$ with the assumption $e_{\rm p}=I$, and taking the volume of the belt to be that of a narrow ring with constant aspect ratio ($V=4\pi R^2\Delta R I$, where $\Delta R$ is the width of the belt in AU), we obtain the general expression
\begin{equation}
R_{\rm col}(D_{\rm min})=0.32\frac{v_{\rm Kep}f}{\Delta R}\left[1+\frac{6q-10}{(3q-4)}+\frac{3q-5}{(3q-3)}\right]=9.6\frac{M_{\star}^{0.5}f}{R^{0.5}\Delta R}\left[1+\frac{6q-10}{(3q-4)}+\frac{3q-5}{(3q-3)}\right],
\end{equation}
which for $q=11/6$, as is the case for the Fomalhaut belt, becomes
\begin{equation}
R_{\rm col}(D_{\rm min})=17.9M_{\star}^{0.5}fR^{-0.5}\Delta R^{-1}.
\end{equation}

This formula differs from Eq. 25 in \citet{Wyatt2007a} and Eq. B4 in \citet{Zuckerman2012}, respectively, by a factor $\sim$1.4. The difference lies in two simplyfing assumptions taken by those authors: 1) the relative collision velocities are set only by the vertical motion of particles in the disk, so that $v_{\rm rel}/v_{\rm Kep}\sim sin(I)\sim I$ rather than $v_{\rm rel}/v_{\rm Kep}=\sqrt{1.25e_{\rm p}^2+I^2}\sim\sqrt{2.25}I$, as assumed here; 2) particles of size $D_{\rm min}$ collide only with particles of the same size $D_{\rm min}$, so that the integral in Eq. \ref{eq:bigFinteg}, corresponding to the expression in square brackets in Eq. \ref{eq:rcolsimpl1}, is independent of the slope of the size distribution and equal to 4, rather than 28/15 (for $q=11/6$). In the Fomalhaut belt, we find that the smallest bodies will collide at a rate of $1.2\times10^{-5}$ yr$^{-1}$, corresponding to a collision timescale of $8.5\times10^4$ yr. This is similar to the timescale derived by \citet{Zuckerman2012}, but almost 2 orders of magnitude longer than the timescale derived by \citet{Acke2012}. We divert the reader to Appendix C in \citet{Zuckerman2012} for a detailed discussion of this discrepancy.

Finally, combining the mass of small grains $M_{D_{\rm min}}$ with their collision rate $R_{\rm col}(D_{\rm min})$ above, we obtain a simple expression for the mass loss rate $\dot{M}_{D_{\rm min}}$ (in M$_{\oplus}$ Myr$^{-1}$) of grains at the bottom of the cascade,
\begin{equation}
\label{eq:mlossratefinal}
\dot{M}_{D_{\rm min}}=1.2\times10^{3} \ R^{1.5}\Delta R^{-1}f^2L_{\star}M_{\star}^{-0.5},
\end{equation}
which is well constrained observationally through SED fitting (yielding stellar properties and the fractional luminosity) together with resolved imaging of the belt (yielding accurate values of the belt radius and width). This allows us to derive a mass loss rate of the smallest grains in the Fomalhaut belt of $1.1\times10^{-2}$ M$_{\oplus}$ Myr$^{-1}$, or $2.1\times10^{12}$ g s$^{-1}$. The latter value is very close to the mass loss rate obtained from the modelling results of \citep{Wyatt2002} for grains $\sim$10 $\mu$m in size ($\sim0.01$ M$_{\oplus}$ Myr$^{-1}$), although differs substantially from the value of $\sim0.1$ M$_{\oplus}$ Myr$^{-1}$ obtained by \citet{Matra2015} derived for the largest bodies participating in the cascade. This unexpected difference in mass loss rate is likely due to the fact that the \citet{Wyatt2002} model assumes a $q=11/6$ size distribution, typical of a collisional cascade where the planetesimal strength ${\rm Q}_{\rm D}^{\star}$ is independent of size, but then uses a ${\rm Q}_{\rm D}^{\star}$ varying with size to calculate collision rates, leading to an inconsistency that causes the unexpected result of a mass loss rate that varies with particle size.

\end{document}